\documentclass[11pt,a4paper]{article}
\usepackage[a4paper,left=2.6cm,right=2.6cm,top=2.8cm,bottom=2.8cm]{geometry}
\usepackage{amsmath}
\IfFileExists{newtxtext.sty}{\usepackage{newtxtext,newtxmath}}{\usepackage{times,amssymb}}
\usepackage[T1]{fontenc}
\usepackage{graphicx}
\graphicspath{{figures/}}
\usepackage{bm}
\usepackage{booktabs,array}
\usepackage[authoryear,round]{natbib}
\usepackage[colorlinks=true,linkcolor=red,citecolor=blue,urlcolor=blue]{hyperref}
\usepackage{orcidlink}
\usepackage{caption}
\newcommand\lefttitle[1]{}
\newcommand\righttitle[1]{}
\newcommand\aff[1]{\textsuperscript{#1}}
\newcommand\affiliation[1]{\date{\normalsize #1}}
\makeatletter\newcommand\corresau[1]{\gdef\@corres{#1}}\gdef\@corres{}\makeatother
\newcommand\email[1]{\href{mailto:#1}{#1}}
\renewcommand\and{and }
\newenvironment{keywords}{\par\vspace{4pt}\noindent\textbf{Key words:} }{\par}
\newenvironment{bmhead}[1][]{\par\vspace{6pt}\noindent\textbf{#1} }{\par}
\renewenvironment{abstract}{\par\vspace{8pt}\noindent\rule{\linewidth}{0.4pt}\par\vspace{4pt}\noindent\textbf{Abstract}\par\vspace{2pt}\noindent\ignorespaces}{\par\vspace{4pt}\noindent\rule{\linewidth}{0.4pt}\par}
\makeatletter
\renewcommand\maketitle{\begin{center}
  {\LARGE\bfseries\@title\par}\vspace{10pt}
  {\large\@author\par}\vspace{6pt}
  {\@date\par}\vspace{4pt}
  {\small Corresponding author: \@corres\par}\end{center}\vspace{6pt}}
\makeatother
\usepackage{etoolbox}
\AtBeginDocument{}
\title{Triadic energy routes in a short laminar separation bubble}

\author{Gustavo Lopes\aff{1}\,\orcidlink{0009-0009-5061-897X}, Pawel Przytarski\aff{2}\,\orcidlink{0000-0003-0104-4606}, Daniele Simoni\aff{1}\,\orcidlink{0000-0003-4161-0721}, \and Davide Lengani\aff{1}\,\orcidlink{0000-0001-6347-0817}}

\affiliation{\aff{1}DIME, University of Genova, Genova, Italy\\
\aff{2}School of Aerospace, Mechanical and Mechatronic Engineering, The University of Sydney, Sydney, Australia}

\corresau{Gustavo Lopes, \email{gustavo.filipe.docarmolopes@edu.unige.it}}

\begin{document}
\maketitle

\begin{abstract}
The kinetic energy budget of a laminar separation bubble on a compressor blade is formed mode by mode in frequency and spanwise wavenumber from a direct numerical simulation at a Reynolds number of $2.5\times10^{5}$ based on the axial chord. The nonlinear transfer between modes is evaluated as a triadic tensor that gives the energy exchanged in every triad, the roles of its three modes, and the location of the exchange. Nine dominant tonal modes are identified: the flapping of the bubble, the Kelvin-Helmholtz mode with its sidebands, a pair of modes near reattachment, and their sum, difference, and harmonic. They hold $8\,\%$ of the fluctuation energy, and their triads are responsible for $19\,\%$ of the energy exchanged. Every tonal mode is fed by the mean shear and exports $69$ to $82\,\%$ of the energy it takes from the mean flow. The Kelvin-Helmholtz mode sends $45\,\%$ of the energy it sheds to lower frequencies. The largest of these transfers is a triad in which it acts as a catalyst. The exchange in each triad is split into nine channels by velocity components. In every triad of spanwise-uniform modes, one channel dominates: the wall-tangential velocity of the donor advected by the wall-normal velocity of the catalyst. For a pair of oblique modes, the same advection by the two-dimensional Kelvin-Helmholtz mode exchanges their spanwise velocity, which the mean flow cannot produce. The rollers thus become three-dimensional through the advecting velocity of the Kelvin-Helmholtz mode, which supplies no energy to the exchange.
\end{abstract}


\section{Introduction}\label{sec:intro}
A laminar boundary layer that separates under an adverse pressure gradient forms a detached shear layer. The shear layer amplifies waves at a selected frequency into spanwise rollers, the rollers break down, and the enhanced mixing reattaches the flow in the mean. The region between separation and reattachment is a laminar separation bubble. The process is separation-induced transition \citep{Gaster1967,Alam2000}. The fluctuations are produced by the mean shear at a few discrete frequencies and at zero spanwise wavenumber, whereas downstream of reattachment their energy is spread across a continuum of frequencies and spanwise scales \citep{Marxen2013,Kurelek2018}. The transfer of energy between these scales is due to the nonlinear terms, which linear stability theory does not retain.

The steps of the breakdown are documented. Under clean inflow, the Kelvin-Helmholtz instability of the inflectional shear layer dominates \citep{Watmuff1999,Rodriguez2010,Hosseinverdi2019}. The spectrum develops a tone at the shedding frequency with companion peaks. Free shear layers show harmonics, subharmonics, and sum and difference frequencies of such tones \citep{Miksad1973,HoHuerre1984}. The same families of peaks appear in separation bubbles, where forcing has traced them to wave interactions and vortex merging \citep{Kurelek2018,Kurelek2019}. The rollers deform along the span \citep{Michelis2018}. Secondary instability breaks them into three-dimensional structures \citep{Alam2000,Jones2008,Marxen2013,Rodriguez2021}. A slow flapping of the bubble modulates the sequence \citep{Rinoie2004,Embacher2014,Fang2024}. In turbulent separation bubbles, this low-frequency motion is a spanwise modulation of the reversed-flow region rather than a two-dimensional contraction \citep{Steinfurth2025}. In laminar bubbles, its low-frequency content contributes to the production of the stresses in the fore part of the bubble \citep{Lengani2017}. Spectra and bicoherence record the peaks and the phase coupling between them \citep{Kurelek2018,Borgmann2025}, but neither measures how much energy moves between the peaks, nor in which direction.

Such measurements exist for other flows. The convective nonlinearity couples Fourier modes in triplets. A triad exchanges energy only if its frequencies and wavenumbers close,
\begin{equation}
  f_l \pm f_m \pm f_n = 0, \qquad \beta_l \pm \beta_m \pm \beta_n = 0,
  \label{eq:closure}
\end{equation}
with $f$ the frequency and $\beta$ the spanwise wavenumber \citep{Waleffe1992,Moffatt2014}. Within a triad, the nonlinearity redistributes energy without creating it. The triadic terms admit an exact spectral representation \citep{Cheung2014}. Spectral budgets of channel flow have mapped forward and inverse cascades \citep{Mizuno2016,Cho2018,Lee2019,Kawata2018}. \citet{Ding2025} quantified the energy transferred between individual pairs of modes. Budgets of coherent kinetic energy assign each member of a triad a role. The donor loses energy, the recipient gains it, and the catalyst enables the exchange without net gain \citep{Baj2017,Biswas2022,Kinjangi2023,Yeung2026}. Bispectral mode decomposition extracts the structures behind the strongest phase couplings \citep{Schmidt2020} and has traced vortex pairing in jets \citep{Nekkanti2025,Nekkanti2025b}. In turbomachinery, triadic transfer rates between data-driven modes have been estimated in a repeating-stage compressor cascade \citep{PrzytarskiCTR2024}, without resolution in frequency and spanwise wavenumber. For separation bubbles, bispectral mode decomposition has identified the dominant frequency triads in a transitional shock-induced bubble \citep{BenHassanSaidi2025} and in a forced flat plate bubble \citep{Borgmann2025}. Both analyses detect which interactions are active, but not the amount or the direction of the energy exchanged. Both are spanwise averaged. The spanwise wavenumber cannot be averaged out of the problem. Resonant triads of oblique waves are the classical route by which two-dimensional waves feed three-dimensional disturbances \citep{Craik1971}. The deformation and breakdown of the rollers are three-dimensional from their onset \citep{Marxen2013,Michelis2018}. A signed budget of the energy transfer, resolved jointly in frequency and spanwise wavenumber and located in space, has not been reported for a laminar separation bubble.

In this work, the kinetic energy budget of every frequency and spanwise-wavenumber mode is formed from a direct numerical simulation of a laminar separation bubble. The nonlinear transfer is evaluated as a triadic tensor that assigns a signed exchange, a donor, a recipient, and a catalyst to every admissible triad and locates each exchange in the bubble. Three questions are addressed, namely where the energy of the tonal modes goes, where in the bubble the exchanges occur, and how many degrees of freedom carry the transfer. The paper is organised as follows. \S\,\ref{sec:methodology} describes the test case, the decomposition, and the tensor formulation. \S\,\ref{sec:results} presents the transfer summed over all modes, the tonal modes and their shapes, their budgets, their triads, the maps of production and transfer, the routes of the tonal energy, and the velocity channels of the routes. \S\,\ref{sec:conclusions} summarises the findings.

\section{Methodology}\label{sec:methodology}

\subsection{Test case}\label{sec:testcase}
\begin{figure}[!h]
  \centering\includegraphics[width=\linewidth]{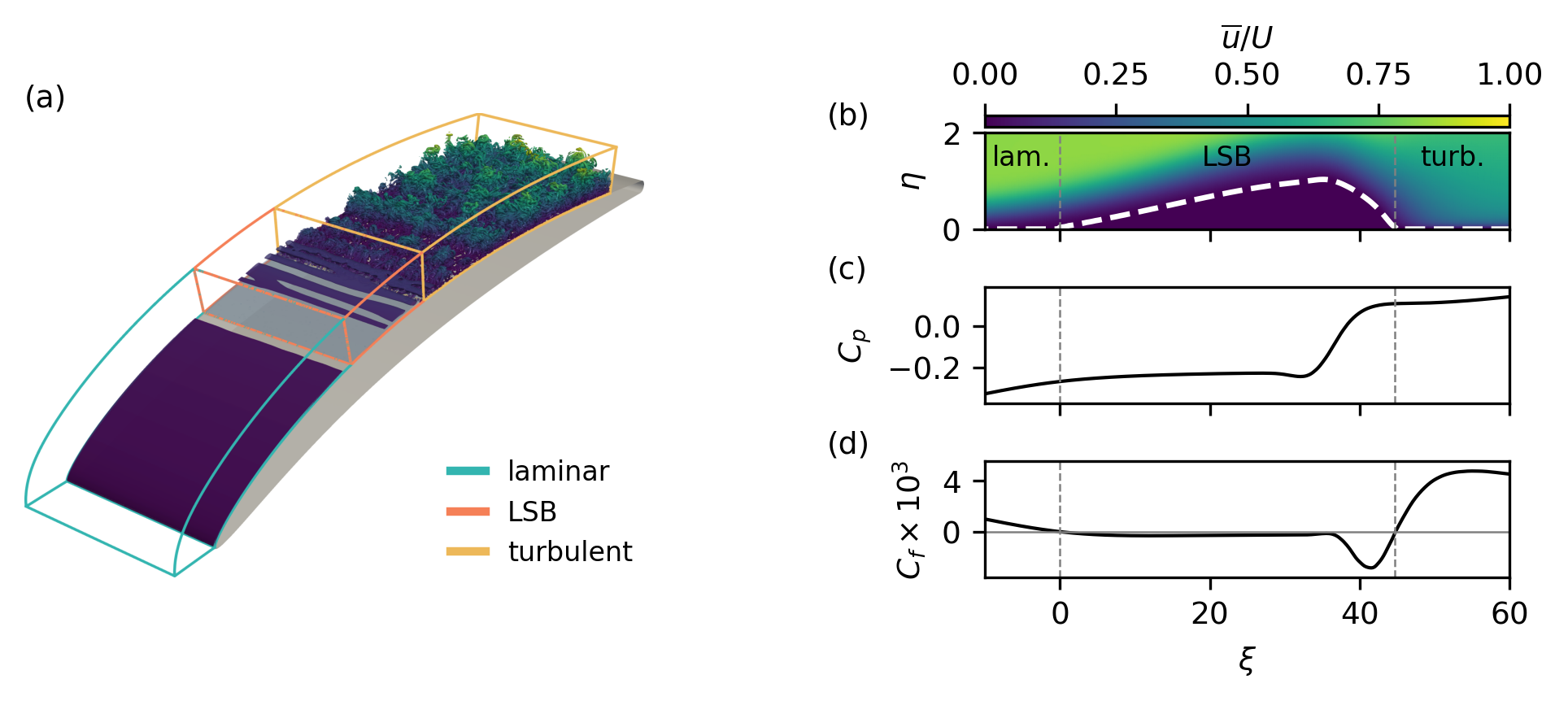}
  \caption{{Overview. (a) Instantaneous isosurfaces of the $Q$-criterion colored by the wall distance. The rollers of the separated shear layer, their breakdown, and the turbulent boundary layer downstream are visible. The wireframe blocks mark the laminar region, the bubble, and the turbulent region. (b) Mean streamwise velocity in the frame $(\xi,\eta)$ of (\ref{eq:frame}), with the $\overline{u}=0$ isoline as a white dashed line. (c) Pressure coefficient and (d) skin-friction coefficient on the suction side against $\xi$. The zero crossings of $C_f$ mark separation, $\xi=0$, and reattachment, $\xi_{r}=44.7$.}}
  \label{fig:overview}
\end{figure}

The flow field is obtained from a direct numerical simulation performed with the high-order compressible solver HiPSTAR, validated for turbomachinery applications \citep{Sandberg2015}. The computational domain is an overset arrangement of a body-fitted O-grid around the blade and a Cartesian background mesh \citep{Deuse2020}. The geometry and the clean laminar inflow follow the controlled-diffusion compressor configuration of \citet{Przytarski2021}. The Reynolds number based on the axial chord and the inlet velocity is $2.5\times10^{5}$. The inlet Mach number is $0.20$. The near-wall O-grid block contains $3000\times200\times256$ points in the streamwise, wall-normal, and spanwise directions, with periodic boundary conditions in the span. In the turbulent boundary layer downstream of reattachment, the spacings along the wall and along the span remain below $\Delta s^{+}=14$ and $\Delta z^{+}=14$, in wall units based on the friction velocity of the span- and time-averaged flow. The first cell is below two wall units except across the skin-friction peak at reattachment, where it reaches $2.5$. Wall units are not a resolution criterion in the bubble, where the friction velocity vanishes at separation and at reattachment. There, the first cell at the wall is $0.23\,\theta_{s}$ high, with $\theta_{s}$ the momentum thickness at separation. Five points lie within $\theta_{s}$ and $30$ within the bubble height $h$. The spacing along the wall is $1.3\,\theta_{s}$, so that the wavelength of the rollers is resolved by $36$ to $56$ points. The spacing along the span is $1.4\,\theta_{s}$, that is $64$ points per quarter of the span. The record analysed comprises $1250$ uniformly sampled snapshots, transformed as a single record without segmentation or windowing. Lengths are normalised by the axial chord $C_{ax}$ and velocities by the inlet velocity $U_{in}$. An instantaneous field, visualised by isosurfaces of the $Q$-criterion coloured by the wall distance in figure~\ref{fig:overview}(a), shows the laminar shear layer, the spanwise rollers, their breakdown and the turbulence downstream of reattachment, together with the streamwise extent of the region analysed. The record resolves $625$ frequencies up to the Nyquist limit. The $256$ spanwise points resolve $128$ spanwise wavenumbers. The frequency resolution is $\Delta St_{\theta}=1.18\times10^{-4}$ and the Nyquist Strouhal number is $St_{\theta}=7.4\times10^{-2}$, where $St_{\theta}=f\theta_{s}/U_{e}$ is built on $\theta_{s}=8.56\times10^{-4}\,C_{ax}$ and the edge velocity at separation, $U_{e}=0.828\,U_{in}$.

A laminar separation bubble forms on the suction side. The mean flow separates at $x/C_{ax}=0.2217$ and reattaches at $x/C_{ax}=0.4320$, from the zero crossings of the skin-friction coefficient and the mean $\overline{u}=0$ isoline. Positions are given in the frame of reference of the bubble. The coordinate along the suction surface is $s$, with $s_{\mathrm{sep}}$ its value at separation, and the wall-normal distance is $d$. Both are normalised by the maximum height $h$ of the mean reversed-flow region,
\begin{equation}
  \xi=\frac{s-s_{\mathrm{sep}}}{h}, \qquad \eta=\frac{d}{h},
  \label{eq:frame}
\end{equation}
so that separation lies at $\xi=0$ and reattachment at $\xi_{r}=44.7$. The spanwise extent is $L_{z}=51.7\,h$. All the maps of the manuscript are presented in this frame. The mean streamwise velocity in this frame, the pressure coefficient, and the skin-friction coefficient on the suction side are given in figure~\ref{fig:overview}(b) to (d). The bubble is shallow, with a height-to-length ratio $h/L_{b}=0.022$, $L_{b}=\xi_{r}h$ being its length along the wall, in the short-bubble range \citep{OwenKlanfer1953,OMeara1987,Diwan2006,Dellacasagrande2023}. Because of the stagger of the blade, velocity components and mean gradients are rotated into the local wall-tangent frame $(s,n,w)$, with $u_s$ the wall-tangential, $u_n$ the wall-normal, and $u_w$ the spanwise component, and $U_s$ and $U_n$ the mean velocity components in this frame. All component-resolved quantities are reported in this frame. 

\begin{figure}[!h]
  \centering\includegraphics[width=\linewidth]{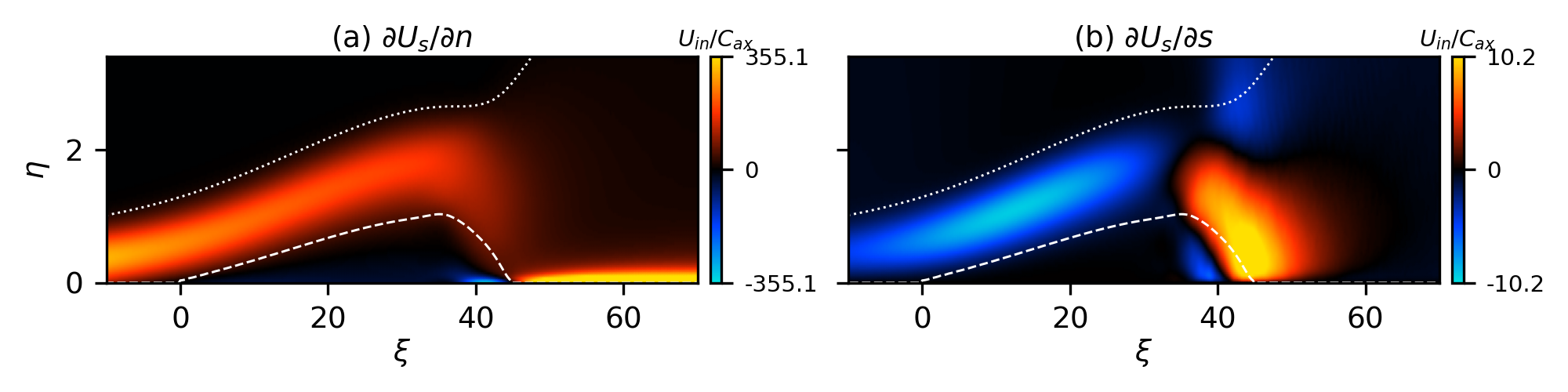}
  \caption{{Mean strains. (a) Mean shear $\partial U_s/\partial n$ and (b) mean streamwise strain $\partial U_s/\partial s$, in units of $U_{in}/C_{ax}$. Dashed line: height of the mean $\overline{u}=0$ isoline. Dotted line: shear-layer edge $\delta_{e}$.}}
  \label{fig:gradients}
\end{figure}

The mean strains in the surroundings of the separated region are given in figure~\ref{fig:gradients}, with the two reference lines that are drawn on every map of the bubble in the following. The dashed line is the height of the mean $\overline{u}=0$ isoline, and the dotted line is the edge of the shear layer, $\delta_{e}$, obtained with the generalised-velocity method of \citet{Coleman2018}. The mean shear $\partial U_s/\partial n$ is largest in the separated shear layer above the reversed-flow region, weakens through reattachment, and is intense again at the wall downstream of reattachment, where a turbulent boundary layer forms. The mean streamwise strain $\partial U_s/\partial s$ is negative along the front part of the bubble. The mean flow decelerates along the wall from the separation point to the maximum height of the bubble and accelerates downstream of it, due to the momentum transfer that leads to reattachment. In a two-dimensional mean flow, these two strains, together with $\partial U_n/\partial s$ and $\partial U_n/\partial n$, are the four strains that produce turbulent kinetic energy. The two shown are the largest.

\subsection{Turbulent kinetic energy transport in scale space}\label{sec:tke}
With $U_i(x,y)$ the span- and time-averaged mean flow, $u_i'$ the fluctuating velocity, and $s_{ij}'=\frac{1}{2}(\partial u_i'/\partial x_j +\partial u_j'/\partial x_i)$, the balance of the fluctuation kinetic energy $k=\frac{1}{2}\overline{u_i'u_i'}$ reads

\begin{multline}
  \underbrace{\frac{\partial k}{\partial t}}_{\mathcal{L}}
  +\underbrace{U_j\frac{\partial k}{\partial x_j}}_{\mathcal{C}}
  =\underbrace{-\overline{u_i'u_j'}\frac{\partial U_i}{\partial x_j}}_{\mathcal{P}}
  \underbrace{-\frac{\partial}{\partial x_j}
    \Big(\frac{1}{2}\overline{u_i'u_i'u_j'}\Big)}_{\mathcal{D}_T}
  \underbrace{-\frac{1}{\rho}\frac{\partial\overline{p'u_j'}}
    {\partial x_j}}_{\mathcal{D}_p}\\
  +\underbrace{2\nu\frac{\partial}{\partial x_j}
    \overline{u_i's_{ij}'}}_{\mathcal{D}_v}
  \underbrace{-2\nu\,\overline{s_{ij}'s_{ij}'}}_{\varepsilon}
  +\underbrace{\frac{1}{\rho}\overline{p'\frac{\partial u_j'}
    {\partial x_j}}}_{\Phi},
  \label{eq:phys}
\end{multline}
with $\rho$ the density, $\nu$ the kinematic viscosity, rate of change $\mathcal{L}$, mean advection $\mathcal{C}$, production $\mathcal{P}$, turbulent transport $\mathcal{D}_T$, pressure transport $\mathcal{D}_p$, viscous diffusion $\mathcal{D}_v$, dissipation $\varepsilon$, and pressure-dilatation $\Phi$. The record is statistically stationary and $\mathcal{L}$ is neglected. The flow is weakly compressible, with $Ma=0.20$, so the pressure-dilatation $\Phi$ is not identically zero. Its integrated value is below $10^{-4}$ of the production for every mode analysed in \S\,\ref{sec:results}. With $\Phi$ neglected and the pressure-strain correlation cancelling in this summed form, the pressure acts only through $\mathcal{D}_p$.

The fluctuation is decomposed by a two-stage unitary discrete Fourier transform, in time over the $N$-snapshot record and in the homogeneous spanwise direction over $N_z$ points,
\begin{equation}
  \hat{u}_i(x,y;\beta,f)=\frac{1}{\sqrt{N N_z}}
  \sum_{q=0}^{N-1}\sum_{r=0}^{N_z-1}
  u_i'(x,y,z_r,t_q)\,
  \mathrm{e}^{-\mathrm{i}2\pi f q/N}\,\mathrm{e}^{-\mathrm{i}2\pi\kappa r/N_z},
  \label{eq:dft}
\end{equation}
with $q$ the snapshot index, $t_q=q\Delta t$ the sampling instants, $r$ the spanwise grid index, $z_r=rL_z/N_z$ the spanwise grid points, $f$ the integer frequency bin, $\kappa$ the integer spanwise mode and $\beta=2\pi\kappa/L_z$ the spanwise wavenumber, so that $2\pi\kappa r/N_z=\beta z_r$. Spanwise wavenumbers are reported as $\beta h$. Throughout, the hat denotes a transformed quantity, an asterisk the complex conjugate, and $\Re$ the real part. The product of two modes at the same $(f,\kappa)$ is complex. Its real part is the sum of the contributions of the mode and of its complex conjugate at $(-f,-\kappa)$. Every term of the modal budget is therefore real, and the modes with $f\geq0$ suffice. The transform is applied first along the span and then in time. Every transformed quantity remains a function of $(x,y)$. All derivatives in $x$ and $y$ are evaluated on the transformed fields in the plane of the blade, whereas the spanwise derivative acts on the transform as a multiplication, $\partial/\partial z\to\mathrm{i}\beta$. The mode $\hat{u}_i(x,y;0,0)$ carries the span- and record-averaged mean, about which the fluctuation is defined. Because the mean is independent of $z$ and of time, every term of the budget that involves the mean, such as the production, is a product of one mode with the mean. Every map shown in this paper is a function of $(x,y)$ at one fixed $(f,\kappa)$.

Transforming the fluctuation momentum equation with (\ref{eq:dft}) and projecting on the mode yields the transport equation of the modal energy $\hat{E}=\frac{1}{2}\hat{u}_i\hat{u}_i^{*}$,
\begin{multline}
  \underbrace{U_j\frac{\partial\hat{E}}{\partial x_j}}_{\hat{\mathcal{C}}}
  =\underbrace{-\Re\big[\hat{u}_i\hat{u}_j^{*}\big]
    \frac{\partial U_i}{\partial x_j}}_{\hat{\mathcal{P}}}
  \underbrace{-\,\Re\Big[\hat{u}_i^{*}\,
    \widehat{u_j'\tfrac{\partial u_i'}{\partial x_j}}\Big]}_{\hat{\mathcal{N}}}
  \underbrace{-\frac{1}{\rho}\frac{\partial}{\partial x_j}
    \Re\big[\hat{p}\,\hat{u}_j^{*}\big]}_{\hat{\mathcal{D}}_p}\\
  +\underbrace{2\nu\frac{\partial}{\partial x_j}
    \Re\big[\hat{u}_i^{*}\hat{s}_{ij}\big]}_{\hat{\mathcal{D}}_v}
  \underbrace{-2\nu\,\hat{s}_{ij}\hat{s}_{ij}^{*}}_{\hat{\varepsilon}},
  \label{eq:spectral}
\end{multline}
with $\hat{s}_{ij}=\frac{1}{2}(\partial\hat{u}_i/\partial x_j +\partial\hat{u}_j/\partial x_i)$. Summing (\ref{eq:spectral}) over all modes recovers (\ref{eq:phys}) without $\Phi$, as shown in appendix~\ref{app:sum}. The cubic nonlinearity enters as the single term $\hat{\mathcal{N}}$, which has no one-to-one counterpart in (\ref{eq:phys}). Mode by mode, it contains both spatial transport and interscale transfer. Only its mode sum reduces to $\mathcal{D}_T$. Integrated budget terms are in units of $U_{in}^{3}C_{ax}^{2}$ and their densities in units of $U_{in}^{3}/C_{ax}$.

With a two-dimensional mean flow, the production comprises four terms, grouped by the acting strain as
\begin{equation}
  \hat{\mathcal{P}}^{s}=-\Re\big[\hat{u}_s\hat{u}_n^{*}\big]
  \Big(\frac{\partial U_s}{\partial n}+\frac{\partial U_n}{\partial s}\Big),
  \qquad
  \hat{\mathcal{P}}^{n}=-\Re\big[\hat{u}_s\hat{u}_s^{*}\big]
  \frac{\partial U_s}{\partial s}
  -\Re\big[\hat{u}_n\hat{u}_n^{*}\big]\frac{\partial U_n}{\partial n},
  \label{eq:prodsplit}
\end{equation}
with $\hat{\mathcal{P}}^{s}$ the shear production, carried by the modal shear stress against the mean shear of figure~\ref{fig:gradients}(a), and $\hat{\mathcal{P}}^{n}$ the normal production, carried by the normal stresses against the mean streamwise strain of figure~\ref{fig:gradients}(b) and its wall-normal counterpart.

Following \citet{Mizuno2016} and \citet{Lee2019}, the nonlinear term is separated into an interscale transfer and a spatial transport by adding and subtracting the divergence of the modal flux $\frac{1}{2}\Re[\hat{u}_i^{*}\widehat{u_i'u_j'}]$,
\begin{equation}
  \hat{\mathcal{N}}=
  \underbrace{-\,\Re\Big[\hat{u}_i^{*}\,
    \widehat{u_j'\tfrac{\partial u_i'}{\partial x_j}}\Big]
  +\frac{1}{2}\frac{\partial}{\partial x_j}
    \Re\Big[\hat{u}_i^{*}\,\widehat{u_i'u_j'}\Big]}_{\hat{\mathcal{T}}}
  \underbrace{-\,\frac{1}{2}\frac{\partial}{\partial x_j}
    \Re\Big[\hat{u}_i^{*}\,\widehat{u_i'u_j'}\Big]}_{\hat{\mathcal{D}}_T},
  \label{eq:split}
\end{equation}
with $\hat{\mathcal{D}}_T$ the turbulent transport of modal energy, whose mode sum is the transport $\mathcal{D}_T$ of (\ref{eq:phys}), and $\hat{\mathcal{T}}$ the interscale transfer. The two parts have different properties. The transport is a divergence, so its volume integral reduces to a boundary flux, and it redistributes energy in space only. The transfer redistributes energy among scales only,
\begin{equation}
  \sum_{f,\beta}\hat{\mathcal{T}}(x,y;\beta,f)=0
  \quad\mbox{pointwise}.
  \label{eq:trzero}
\end{equation}

\subsection{Triadic transfer tensor and transfer densities}\label{sec:tensor}
Under the transform (\ref{eq:dft}), the nonlinear term $\hat{\mathcal{N}}$ of (\ref{eq:spectral}) is a discrete convolution over frequency and spanwise wavenumber. Expanding the convolution resolves $\hat{\mathcal{N}}$ into triads. A donor mode $l$ and a catalyst mode $m$ deliver energy to the recipient mode $n$, with $f_n=f_l+f_m$ and $\kappa_n=\kappa_l+\kappa_m$, the one-sided form of the closure condition (\ref{eq:closure}). The catalyst is the advecting mode. In what follows, $l$, $m$, and $n$ denote frequency bins. The energy transferred to the mode $n$ by the pair $(l,m)$ is
\begin{equation}
  \mathcal{T}[l,m]=-\int_V\Re\!\left[\hat{u}_i^{\,n*}\,\hat{u}_j^{\,m}\,
  \frac{\partial\hat{u}_i^{\,l}}{\partial x_j}\right]\mathrm{d}V,
  \label{eq:tensor}
\end{equation}
positive when the recipient gains, where $\hat{u}^{\,l}$ denotes the transform (\ref{eq:dft}) in time only, a function of $(x,y,z)$. The integral over the span sums every spanwise triad $(\kappa_l,\kappa_m)\to\kappa_l+\kappa_m$ of the three frequencies. Equation (\ref{eq:tensor}) is thus the exchange between frequencies with the spanwise modes summed. When a single spanwise triad is needed, in \S\,\ref{sec:transfer} and \S\,\ref{sec:spatial}, the modes of (\ref{eq:dft}) at three fixed spanwise wavenumbers replace the time-only transforms in (\ref{eq:tensor}), and the spanwise mode of each member is stated. The notation $\mathcal{T}[l,m]$ names the donor and the catalyst only, the recipient being fixed by the closure, $n=l+m$. The transport part of $\hat{\mathcal{N}}$ integrates to a boundary flux, so (\ref{eq:tensor}) equals the volume integral of the transfer $\hat{\mathcal{T}}$ of (\ref{eq:split}) and is evaluated on the full nonlinear term. The split is needed only for the pointwise budgets \citep{Waleffe1992,Cheung2014}. The tensor is evaluated from the full record of $N$ snapshots and $N_z$ spanwise points, for every frequency pair and every spanwise mode, with the GPU-accelerated framework of \citet{Lopes2026}, on the LEONARDO supercomputer at CINECA. The donor sum reconstructs the nonlinear term of each recipient, $\hat{\mathcal{N}}(n)=\sum_l \mathcal{T}[l,n-l]$.

The entry $\mathcal{T}[l,m]$ is the energy that the recipient $n$ receives from the donor $l$ through the catalyst $m$. Its partner is $\mathcal{T}[n,-m]$, with $n$ as donor, $l$ as recipient, and the conjugate of the catalyst as advecting mode. It is the energy that $l$ receives from $n$ through the same catalyst. Their sum is a boundary flux of the transport. Their difference is the energy exchanged between $l$ and $n$. It is conserved over the triad \citep{Waleffe1992,Cheung2014,Yeung2026}. Only this antisymmetric part,
\begin{equation}
  A[l,m]=\tfrac{1}{2}\big(\mathcal{T}[l,m]-\mathcal{T}[n,-m]\big),\qquad n=l+m,
  \label{eq:antisym}
\end{equation}
is used in what follows. The present evaluation satisfies its conservation to machine precision. Three modes $l$, $m$, and $n=l+m$ and their complex conjugates $-l$, $-m$, and $-n$ form several entries of $A$, whose set is referred to as a triad. Within a triad, the roles follow \citet{Yeung2026}. The net gain of one of its modes, denoted $k$ because it may be any of $l$, $m$, and $n$, is the sum of the entries of the triad in which $k$ is the recipient,
\begin{equation}
  G_{k}=\sum_{\substack{(l',m')\in\text{triad}\\ l'+m'=k}} A[l',m'],
  \label{eq:gain}
\end{equation}
so that $G_k>0$ when $k$ receives more from the two other modes than it gives them. The efficiency
\begin{equation}
  \gamma_{k}=\frac{G_{k}}{\sum_{(l',m')\in\text{triad}}|A[l',m']|}
  \label{eq:eff}
\end{equation}
divides the net gain by the sum of the absolute values of all entries of the triad. The donor has $\gamma<0$ and the recipient $\gamma>0$. The catalyst is the member whose $|\gamma|$ is an order of magnitude below the others \citep{Baj2017}.

The spatial structure of each interaction is the pointwise transfer density,
\begin{equation}
  \tau(x,y)=-\sum_{i}\sum_{j}\Re\!\left[\hat{u}_i^{\,n*}\,\hat{u}_j^{\,m}\,
  \frac{\partial\hat{u}_i^{\,l}}{\partial x_j}\right],
  \label{eq:tau}
\end{equation}
the integrand of (\ref{eq:tensor}) with the sums over the velocity components written out. Positive $\tau$ marks where the recipient gains. With the time-transformed fields, $\tau$ is a function of $(x,y,z)$ and contains every spanwise triad of the three frequencies. With the modes of (\ref{eq:dft}) at fixed spanwise wavenumbers, it is a function of $(x,y)$ and is the density of one spanwise triad. Summed over all donors $l$ of a recipient $n$, $\tau$ gives the pointwise nonlinear term $\hat{\mathcal{N}}(n)$ of (\ref{eq:spectral}), and, up to the transport, its transfer $\hat{\mathcal{T}}(n)$, so that $\tau$ is the contribution of one triad to $\hat{\mathcal{T}}$.

Following the donor-catalyst component split introduced for channel flow by \citet{DeSalisYoung2024} and \citet{ChenGarciaMayoral2026}, each term of the double sum in (\ref{eq:tau}) is kept separately,
\begin{equation}
  \tau_{ij}=-\Re\!\left[\hat{u}_i^{\,n*}\,\hat{u}_j^{\,m}\,
  \frac{\partial\hat{u}_i^{\,l}}{\partial x_j}\right],
  \qquad \tau=\sum_{i}\sum_{j}\tau_{ij},
  \qquad i,j\in\{s,n,w\},
  \label{eq:tau9}
\end{equation}
the nine channels, with $i$ the recipient component, $j$ the advecting component of the catalyst, and $\partial/\partial w=\mathrm{i}\beta_l$. A single channel is not a tensor component, so its share of an exchange depends on the frame in which the components are taken. In summary, $\tau_{ij}$ is one channel of one triad, $\tau$ the sum of its nine channels, $\mathcal{T}[l,m]$ the volume integral of $\tau$, and $\hat{\mathcal{N}}(n)$ and $\hat{\mathcal{T}}(n)$ the sums over the triads of the mode $n$.

\section{Results and discussion}\label{sec:results}
The results are presented from the whole spectrum to single triads. The production and the transfer summed over all modes identify the frequencies at which both act (\S\,\ref{sec:reduced}). The strongest exchanges select nine tonal modes, identified with the structures of the bubble by their shapes (\S\,\ref{sec:tonalset}). The budget of each gives its production and its transfer (\S\,\ref{sec:budgets}). The triads responsible for the transfer are then ranked (\S\,\ref{sec:selection}). The maps then locate the production and the transfer in the bubble (\S\,\ref{sec:maps}), the routes of the tonal energy are followed triad by triad (\S\,\ref{sec:transfer}), and each route is split into velocity channels (\S\,\ref{sec:spatial}).

\subsection{The transfer summed over all modes}\label{sec:reduced}
\begin{figure}[!h]
  \centering\includegraphics[width=\linewidth]{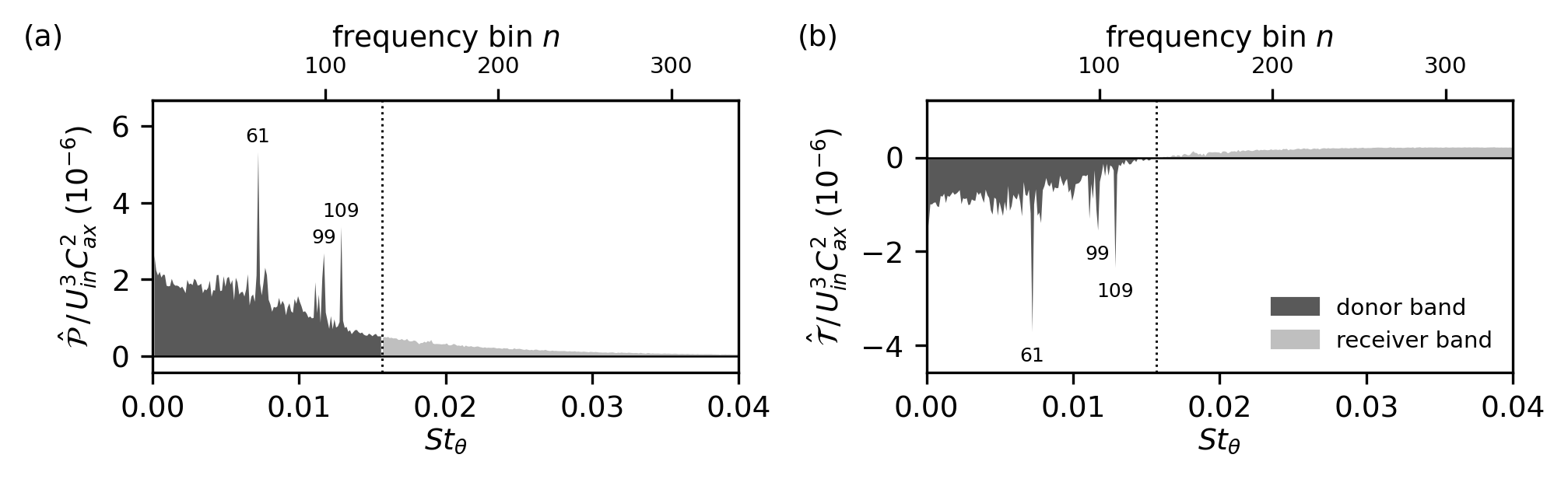}
  \caption{{(a) Production and (b) net interscale transfer of every frequency mode, summed over the span, in $U_{in}^{3}C_{ax}^{2}$. Dark: the donor band of the transfer. Light: its receiver band. Dotted line: the sign change of the transfer. Numbers: the frequency bins of the strongest peaks.}}
  \label{fig:synthesis}
\end{figure}

The production and the net interscale transfer $\sum_m A[l,m]$ of every frequency mode $l$, summed over the spanwise modes, are given in figure~\ref{fig:synthesis}. The transfer changes sign once along the frequency axis, at $St_{\theta}=1.57\times10^{-2}$, which is $2.2$ times the frequency of the Kelvin-Helmholtz mode identified in \S\,\ref{sec:tonalset}. Every mode below this frequency is a net donor and every mode above a net receiver. The donor band contains $21\,\%$ of the resolved frequencies, $70\,\%$ of the fluctuation energy and $82\,\%$ of the production. The production spectrum consists of a broadband base with three peaks, at the Kelvin-Helmholtz frequency, bin $61$, and at the bins $99$ and $109$, tagged in the figure. A single shear-layer frequency is not the rule for a bubble. A converged two-dimensional simulation shows a band of frequencies around the most amplified one of linear theory rather than a single peak \citep{Rist2002}. A self-excited bubble on a wavy surface shows two Kelvin-Helmholtz frequencies and a peak at their difference \citep{Moniripiri2025}. Here the modes $61$ and $109$ differ by the mode $48$, which is not a peak of the spectrum but enters the strongest triads of \S\,\ref{sec:tonalset}. The transfer spectrum has a peak of the opposite sign at each of the three. Its remaining peaks, at the bins $66$, $94$, $96$ and $118$, are weaker, and the first three are neighbours of the modes $65$ and $98$ that enter the ranked triads of \S\,\ref{sec:selection}. The production and the transfer, the linear and the nonlinear terms of (\ref{eq:spectral}), are therefore active at the same frequencies. In the donor band, the transfer removes about half of the production in every bin, with a median of $45\,\%$ against $69$ to $82\,\%$ for the tonal modes of \S\,\ref{sec:budgets}, so that the tonal modes export a larger share of their production than the broadband. In the receiver band, the production decays to zero and the transfer is the only source. The transport terms are below $2\,\%$ of the production and the mean advection is $9\,\%$. The mean shear thus feeds a band of low frequencies and the transfer moves their energy to a continuum of higher frequencies. The net transfer of a mode, however, is the sum of its exchanges with all other modes. It does not identify the partners of an exchange, its direction between two given modes, or its location in the bubble. The following sections resolve it into triads.

\subsection{The tonal set}\label{sec:tonalset}
\begin{figure}[!h]
  \centering\includegraphics[width=\linewidth]{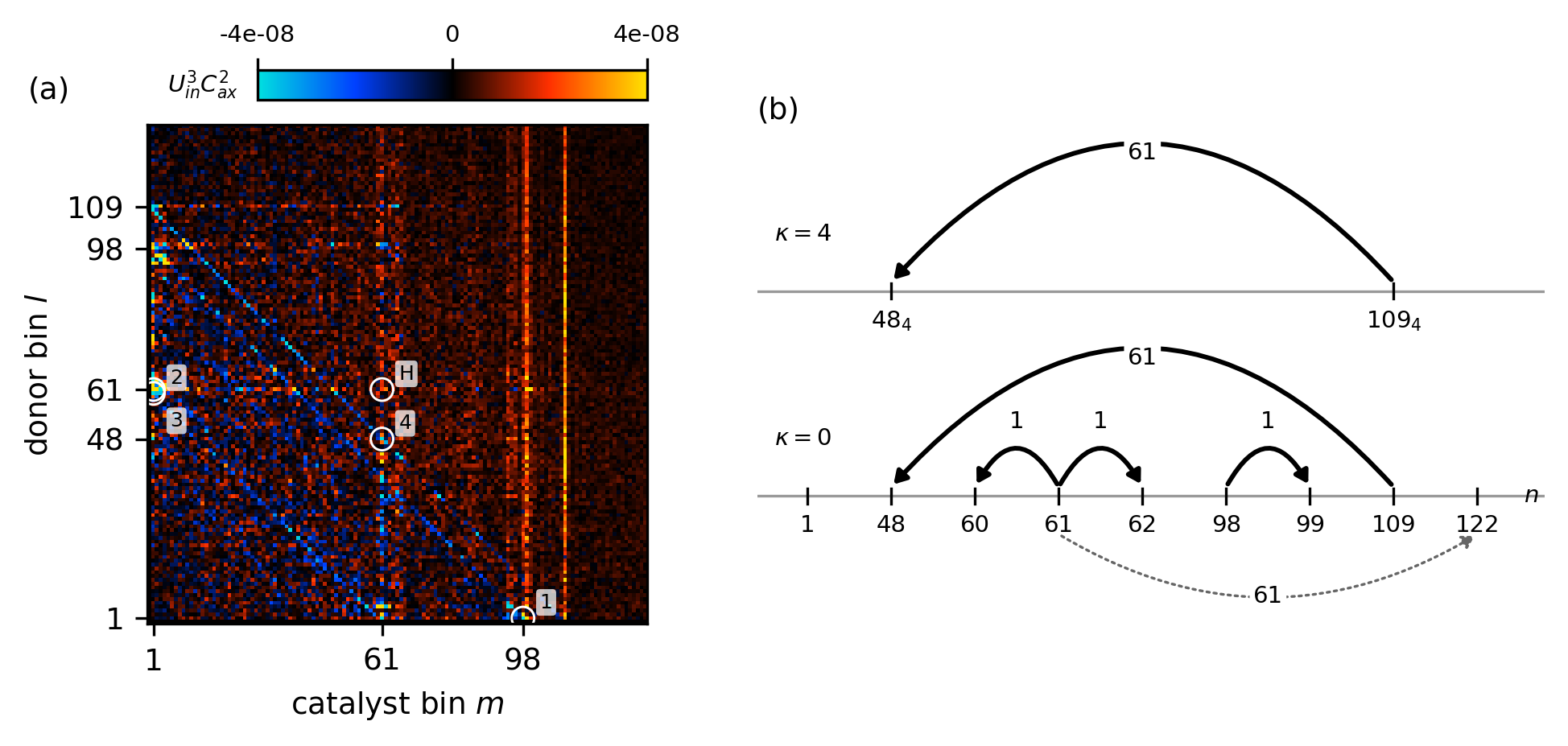}
  \caption{{Mode and triad selection. (a) The exchange matrix $A[l,m]$ of (\ref{eq:antisym}), summed over the spanwise modes, for donor $l$ and catalyst $m$, recipient $n=l+m$, positive when the recipient gains. Circles mark the triads of the tonal set. Tags give their rank in table~\ref{tab:triads}, with H the harmonic triad. (b) The routes of the tonal set. Modes in frequency order on the planes $\kappa=0$ and $\kappa=4$. Arrows from donor to recipient, labelled by the third member of the triad. Dotted: the harmonic triad.}}
  \label{fig:selection}
\end{figure}

The exchange matrix $A$ of (\ref{eq:antisym}) identifies the partners of each exchange (figure~\ref{fig:selection}a). Each entry $(l,m)$ is the net energy transferred to the recipient $n=l+m$ from the donor $l$ with the catalyst $m$, summed over every spanwise triad of the three frequencies. Positive entries mark a gain of the recipient. The notation $(l,m)\to n$ used in the tables lists the two modes whose product feeds the mode $n$. A negative entry means that $n$ loses energy to $l$, so that $n$ is the donor and $l$ the recipient of that triad. Modes are labelled by their frequency bin $n$, with $St_{\theta}=n\,\Delta St_{\theta}$. Four features organise the matrix. The column $m=1$ is the flapping of the bubble, at $St_{\theta}=1.2\times10^{-4}$. The entries of this column are the exchanges catalysed by the flapping mode. The strongest of them, marked by circles in figure~\ref{fig:selection}(a), involve the Kelvin-Helmholtz mode $n=61$ at $St_{\theta}=7.2\times10^{-3}$ and its two neighbours $n=60$ and $62$, the mode $n=109$ and its neighbours, and the two modes $n=98$ and $99$ at $St_{\theta}=11.6$ and $11.7\times10^{-3}$. The columns $m=61$, $98$ and $99$ contain the exchanges catalysed by the Kelvin-Helmholtz mode and by the modes $98$ and $99$. The entry $(l,m)=(48,61)$ is the exchange between the modes $48$ and $109=61+48$ catalysed by the Kelvin-Helmholtz mode. The entry $(61,61)$, marked H, is the self-interaction of the Kelvin-Helmholtz mode that produces its harmonic $n=122$. The nine modes that these entries connect, $n\in\{1,48,60,61,62,98,99,109,122\}$, are referred to as the tonal set. Only three of them, the modes $61$, $99$ and $109$, are peaks of the production spectrum of figure~\ref{fig:synthesis}. The others are identified by the exchanges they take part in. The entries form the routes sketched in figure~\ref{fig:selection}(b) and analysed in \S\,\ref{sec:transfer}. The oblique triad of the sketch is the counterpart of the entry $(48,61)$ on the plane $\kappa=4$, with the two-dimensional Kelvin-Helmholtz mode as catalyst and the modes $48$ and $109$ at $\kappa=4$ as donor and recipient. It is the triad through which the Kelvin-Helmholtz mode couples the plane $\kappa=0$ to the oblique plane. It is examined in \S\,\ref{sec:spatial} for that reason.

\begin{figure}[!h]
  \centering\includegraphics[width=\linewidth]{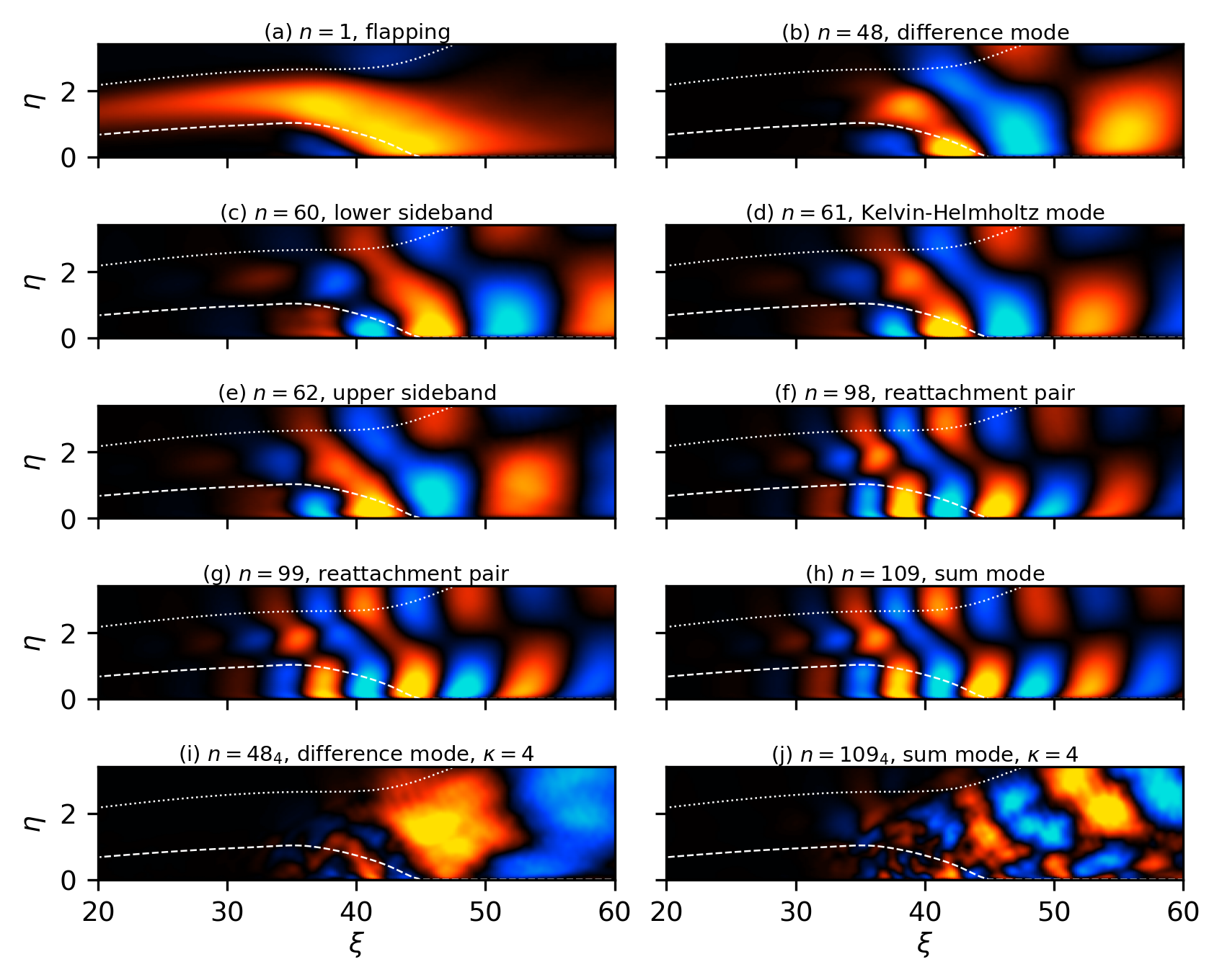}
  \caption{The tonal modes: wall-tangential velocity $\Re[\hat{u}_s]$ of each mode of the tonal set at $\kappa=0$, (a) to (h), and of the oblique difference and sum modes at $\kappa=4$, (i) and (j). Each panel is scaled by its own maximum. Dashed line: $\overline{u}=0$ height. Dotted line: $\delta_{e}$.}
  \label{fig:modes}
\end{figure}

The shapes of the modes of the tonal set identify them with the known structures of the bubble (figure~\ref{fig:modes}). The mode $n=1$ is a single structure that follows the shear layer along the whole bubble and is the flapping of the bubble, the low-frequency motion of separated shear layers \citep{Rinoie2004,Embacher2014,Lengani2017,Fang2024}. The mode $n=61$ is a wave train that grows along the shear layer from $\xi=30$ and reaches the wall downstream of $\xi=42$, the Kelvin-Helmholtz mode of the inflectional profile \citep{Watmuff1999,Rodriguez2010,Marxen2013}. The modes $60$ and $62$ are its sidebands, wave trains of nearly the same wavelength. The modes $98$ and $99$ are wave trains of shorter wavelength located near reattachment and are called the reattachment pair. The mode $109$ has the sum of the frequencies of the modes $61$ and $48$. The mode $48$ has their difference, the frequency of the mode $109$ minus that of the mode $61$. The mode $109$ is a short-wavelength train at the transition front. The mode $48$ is a long-wavelength structure that fills the reattached region. The oblique modes $48_4$ and $109_4$ are the spanwise-periodic counterparts of the last two at $\kappa=4$, the wavenumber retained in \S\,\ref{sec:selection}. They are the members of the oblique triad of figure~\ref{fig:selection}(b).

\subsection{Per-mode budgets of the tonal set}\label{sec:budgets}
\begin{figure}[!h]
  \centering\includegraphics[width=\linewidth]{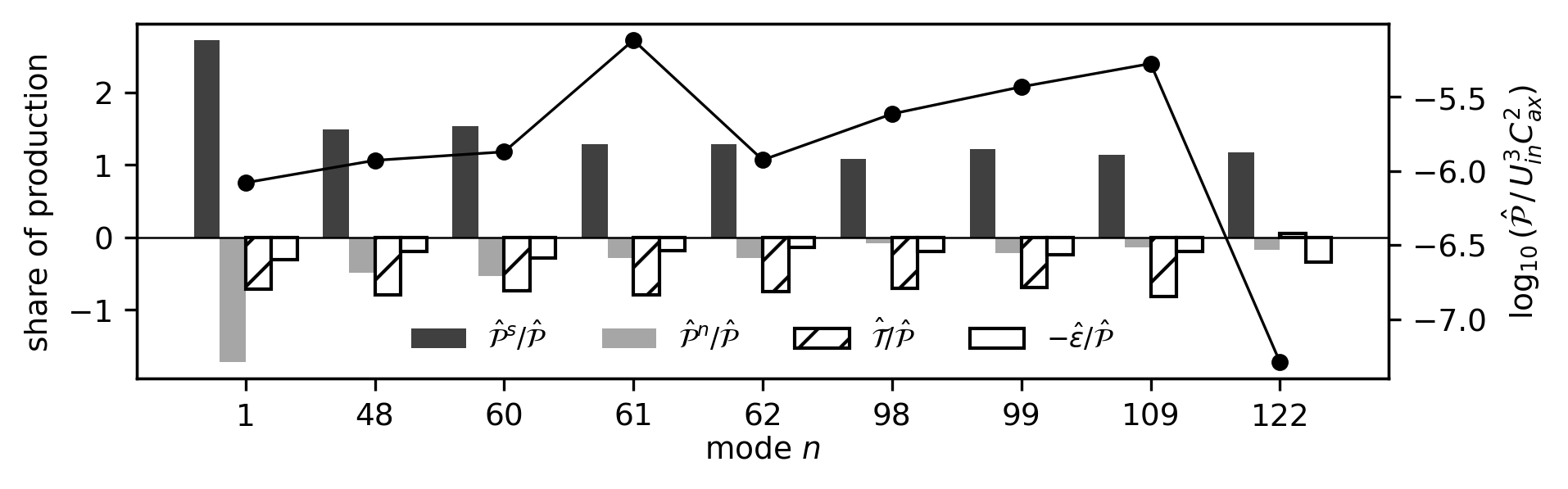}
  \caption{Integrated budget of the nine tonal modes at $\kappa=0$: shear production, normal production, transfer, and dissipation of each mode divided by its production, so that the two production bars sum to one, with the production itself as dots on the right axis.}
  \label{fig:composition}
\end{figure}

Every tonal mode is fed by the mean shear and exports most of what it receives. The budget of each mode of the set (table~\ref{tab:budgets}) is obtained by integrating each term of (\ref{eq:spectral}) in the domain shown in the maps and dividing it by the production of the mode. The transfer removes $69$ to $82\,\%$ of the production of every mode. The dissipation takes the rest. The transport terms are between $0.2\,\%$ and $3\,\%$ of the production for every mode. \citet{Biswas2022} classify a mode as primary if it is fed by production and as secondary if it is fed by the transfer. By this classification, all modes are primary except the harmonic $n=122$, which receives a weak net transfer, $\hat{\mathcal{T}}/\hat{\mathcal{P}}=+0.05$, and is dominated by dissipation. It is labelled mixed in table~\ref{tab:budgets}. The same shares are drawn as bars in figure~\ref{fig:composition}, with the production itself as dots. The shear and normal bars sum to one. The production spans two orders of magnitude across the set. The transport terms are not drawn.

\begin{table}
  \centering
  \caption{Integrated budget of each mode of the tonal set at $\kappa=0$, as shares of the production of the mode. Class after \citet{Biswas2022}: primary when the mode is fed by production and drained by the transfer, mixed when the transfer also feeds it.}
  \label{tab:budgets}
  \begin{tabular}{rrrrrl}
  \toprule
  $n$ & $\hat{\mathcal{P}}^{s}/\hat{\mathcal{P}}$ & $\hat{\mathcal{P}}^{n}/\hat{\mathcal{P}}$ & $\hat{\mathcal{T}}/\hat{\mathcal{P}}$ & $\hat{\varepsilon}/\hat{\mathcal{P}}$ & class \\
  \midrule
  $1$   & $2.72$ & $-1.72$ & $-0.71$ & $0.31$ & primary \\
  $48$  & $1.49$ & $-0.49$ & $-0.79$ & $0.19$ & primary \\
  $60$  & $1.53$ & $-0.53$ & $-0.73$ & $0.29$ & primary \\
  $61$  & $1.28$ & $-0.28$ & $-0.79$ & $0.19$ & primary \\
  $62$  & $1.29$ & $-0.29$ & $-0.75$ & $0.14$ & primary \\
  $98$  & $1.09$ & $-0.09$ & $-0.71$ & $0.20$ & primary \\
  $99$  & $1.22$ & $-0.22$ & $-0.69$ & $0.24$ & primary \\
  $109$ & $1.14$ & $-0.14$ & $-0.82$ & $0.20$ & primary \\
  $122$ & $1.18$ & $-0.18$ & $+0.05$ & $0.34$ & mixed \\
  \bottomrule
  \end{tabular}
\end{table}

The production has a sign structure that the component split exposes. The shear production exceeds the total for every mode. The normal production returns $8$ to $35\,\%$ of the shear production to the decelerating mean, and $63\,\%$ for the flapping mode. Negative production by the normal strains downstream of separation is known from separating and reattaching flows \citep{Cimarelli2019}, and the mode-resolved budget shows it to hold for every tonal mode. The sign of the normal production follows from (\ref{eq:prodsplit}). For a two-dimensional mean, continuity gives $\partial U_n/\partial n=-\partial U_s/\partial s$, and the normal production reduces to
\begin{equation}
  \hat{\mathcal{P}}^{n}=-\big(|\hat{u}_s|^{2}-|\hat{u}_n|^{2}\big)\frac{\partial U_s}{\partial s}.
  \label{eq:pn}
\end{equation}
The mean decelerates along the front part of the bubble (figure~\ref{fig:gradients}b), so a net extraction means that $|\hat{u}_n|$ exceeds $|\hat{u}_s|$ in the production cell, as expected for rollers, whose wall-normal velocity dominates in the cores \citep{Marxen2013}. The sign separates the present flapping mode from the flapping motion of a geometry-induced separation, in which the same strain produces energy because the flapping is a streamwise-elongated structure of wall-tangential velocity, with $|\hat{u}_s|>|\hat{u}_n|$, sustained by a lift-up mechanism \citep{Fang2024}. A low-frequency mode that returns energy to the mean flow through the normal strains has also been found in the POD budget of a flat plate bubble \citep{Lengani2017}.

\subsection{Dominant triads}\label{sec:selection}
The strongest entries of $A$, ranked by the magnitude of $A[l,m]$ after summing the entries that belong to the same three modes, are listed in table~\ref{tab:triads}. The ranking is by frequency. The spanwise triads of a frequency triad are not ranked separately. For the backward triad, the spanwise-uniform triad with the three members at $\kappa=0$ carries $63\,\%$ of the entry, and the triad with the oblique pair at $\kappa=4$ carries $3.4\,\%$ (\S\,\ref{sec:spatial}). All nine modes of the tonal set take part in the ranked triads. The difference mode $48$ is a case in point. Its production is $15\,\%$ of that of the Kelvin-Helmholtz mode, and it is not a peak of either spectrum, yet it is the recipient of the backward triad, the fourth strongest exchange of table~\ref{tab:triads}. The harmonic triad $(61,61)\to122$ does not enter the ranking. None of the tonal frequencies is imposed on the bubble from outside. In the laminar boundary layer upstream of separation, the Kelvin-Helmholtz frequency and the sum mode stand above the broadband level at an energy three to six orders of magnitude below that in the bubble, and the others are at the broadband level. All grow by three to six orders of magnitude through the bubble. The pressure side boundary layer of the adjacent blade and the trailing edge show no peak at any of these frequencies.

\begin{table}
  \centering
  \caption{Ranked triads of the exchange matrix $A$, summed over the spanwise modes, with the entries of the same three modes and their complex conjugates summed. A negative index denotes the complex-conjugate mode, $-m$ standing for $\hat{u}(-f_m)=\hat{u}(f_m)^{*}$, which is why such entries do not appear in the plane of figure~\ref{fig:selection}(a). Positive: the mode $n$ gains from $l$. Negative: the mode $n$ loses to $l$.}
  \label{tab:triads}
  \begin{tabular}{l>{\raggedright\arraybackslash}p{6.4cm}r}
  \toprule
  $(l,m)\to n$ & identification & exchange ($10^{-7}\,U_{in}^{3}C_{ax}^{2}$) \\
  \midrule
  $(98,1)\to99$    & reattachment pair, flapping third member & $+4.55$ \\
  $(60,1)\to61$    & lower sideband of the Kelvin-Helmholtz mode & $-2.49$ \\
  $(62,-1)\to61$   & upper sideband of the Kelvin-Helmholtz mode & $-1.68$ \\
  $(48,61)\to109$  & backward triad, Kelvin-Helmholtz catalyst & $-1.64$ \\
  $(109,-1)\to108$ & lower sideband of the sum mode & $+1.49$ \\
  $(65,1)\to66$    & sideband of $n=65$, broadband neighbour of $n=61$ & $-1.43$ \\
  $(109,10)\to99$  & sum-to-reattachment difference & $-1.38$ \\
  $(64,1)\to65$    & sideband of $n=65$, broadband neighbour of $n=61$ & $-1.36$ \\
  $(110,-1)\to109$ & upper sideband of the sum mode & $+1.22$ \\
  $(94,4)\to98$    & reattachment difference partner & $+1.19$ \\
  $(4,61)\to65$    & scattering of the Kelvin-Helmholtz mode & $+1.17$ \\
  \bottomrule
  \end{tabular}
\end{table}

\begin{figure}[!h]
  \centering\includegraphics[width=\linewidth]{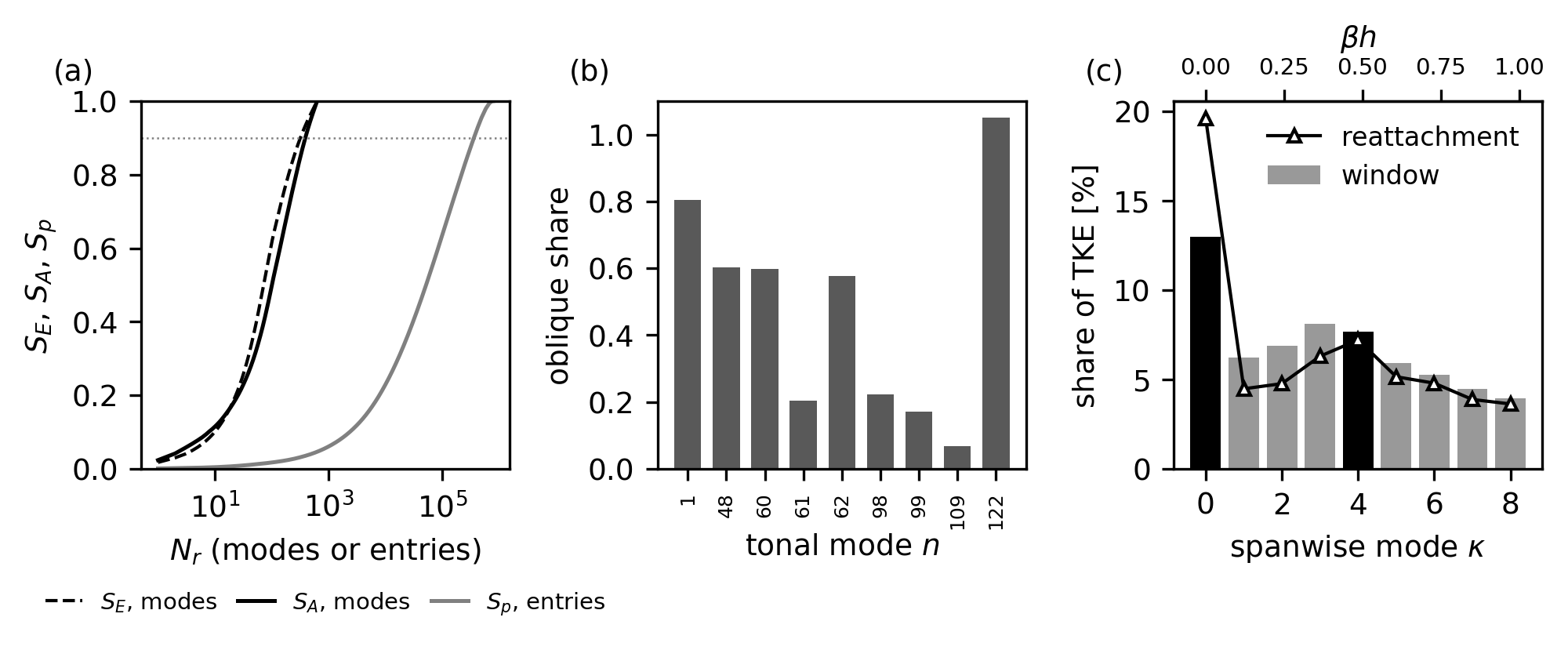}
  \caption{{Share of the exchange carried by the tonal set. (a) Cumulative shares $S_E$, $S_A$ and $S_p$ of (\ref{eq:covE}) to (\ref{eq:covp}) against the number $N_r$ of ranked modes or entries. The dotted line marks $90\,\%$. (b) Share of the net transfer of each tonal mode delivered to its oblique components $\kappa\neq0$. (c) Share of the fluctuation energy per spanwise mode $\kappa$, with the wavenumbers $\pm\beta$ summed, in the whole window and in the region $35<\xi<55$ near reattachment (triangles). Black: the two modes shown in figure~\ref{fig:spanshapes}.}}
  \label{fig:coverage}
\end{figure}

The tonal set holds $8.2\,\%$ of the fluctuation energy. The entries of $A$ that involve at least one mode of the set carry $18.6\,\%$ of the sum of $|A|$ over all entries. The entries between two modes of the set carry $1.2\,\%$. How concentrated the exchange is follows from three cumulative shares (figure~\ref{fig:coverage}a). The first two rank the modes by decreasing energy $\hat{E}_l$ and by decreasing exchange $\sum_m|A[l,m]|$, and give the share of the energy and of the exchange carried by the first $N_r$ modes of each ranking. The third ranks the individual entries of $A$, that is, the donor-catalyst pairs, by decreasing $|A[l,m]|$, and gives the share of the exchange carried by the first $N_r$ entries. With $\mathcal{M}_E(N_r)$ the set of the $N_r$ modes of largest energy, $\mathcal{M}_A(N_r)$ the set of the $N_r$ modes of largest exchange $\sum_m|A[l,m]|$, and $\mathcal{Q}(N_r)$ the set of the $N_r$ entries $(l,m)$ of largest $|A[l,m]|$, the three shares are
\begin{align}
  S_E(N_r)&=\frac{\sum_{l\in\mathcal{M}_E(N_r)}\hat{E}_l}{\sum_l\hat{E}_l},
  \label{eq:covE}\\
  S_A(N_r)&=\frac{\sum_{l\in\mathcal{M}_A(N_r)}\sum_m|A[l,m]|}{\sum_{l,m}|A[l,m]|},
  \label{eq:covA}\\
  S_p(N_r)&=\frac{\sum_{(l,m)\in\mathcal{Q}(N_r)}|A[l,m]|}{\sum_{l,m}|A[l,m]|}.
  \label{eq:covp}
\end{align}
Half of the exchange is carried by about $10^{2}$ modes, or by $5\times10^{4}$ entries, which is $7\,\%$ of the entries of $A$. Ninety per cent is carried by $389$ modes or by $3.6\times10^{5}$ entries, $46\,\%$ of $A$, against $319$ modes for $S_E=0.9$. Ninety-nine per cent requires $82\,\%$ of the entries. The curves $S_E$ and $S_A$ are close to each other and $S_p$ lies below them. The share of the net transfer of each tonal mode that is delivered to its oblique components, $\kappa\neq0$, follows from the transfer resolved in $(f,\kappa)$ (figure~\ref{fig:coverage}b). It is $20\,\%$ for the Kelvin-Helmholtz mode, $7$ to $22\,\%$ for the sum mode and the reattachment pair, $58$ to $60\,\%$ for the sidebands and the difference mode, and $80\,\%$ for the flapping mode. For the harmonic, the net transfer at $\kappa=0$ is close to zero, and the share exceeds one. A description restricted to the plane $\kappa=0$ therefore retains $80\,\%$ of the transfer of the Kelvin-Helmholtz mode and $20\,\%$ of that of the flapping mode. The flapping mode is spanwise uniform at $\kappa=0$ and delivers most of its transfer to oblique components. This is in line with the spanwise-modulated low-frequency motion of separation bubbles observed in volumetric measurements \citep{Steinfurth2025}. Most of the exchange is carried neither by a small number of modes nor by a small number of donor-catalyst pairs. Resolvent analysis of a separation bubble finds the linear amplification to be low-rank, with the Kelvin-Helmholtz mechanism as the leading response \citep{Yeh2020}. The production of figure~\ref{fig:synthesis} has three peaks, not one, and the nonlinear forcing that closes the budget is not low-rank. The tonal set is analysed because it contains the strongest individual triads. These describe where the tonal energy goes and through which velocity component. They do not carry the exchange.

\begin{figure}[!h]
  \centering\includegraphics[width=\linewidth]{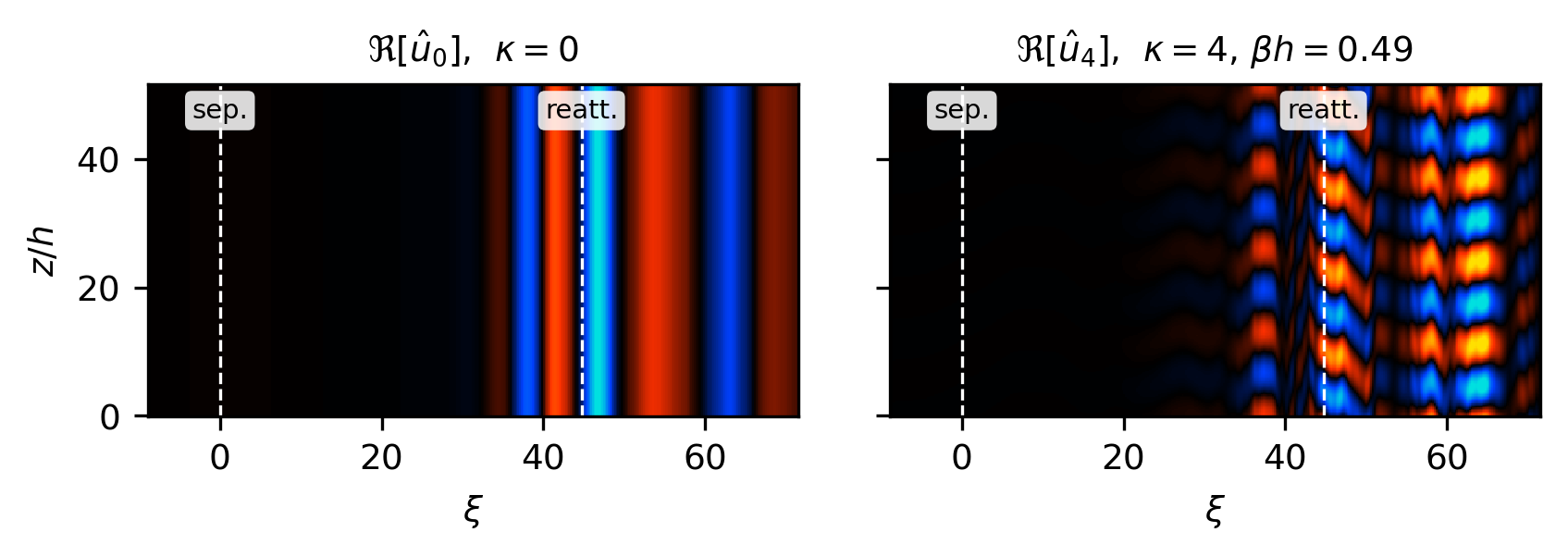}
  \caption{{Wall-parallel shapes of the wall-tangential velocity $\Re[\hat{u}_{s}]$ of the two spanwise modes analysed, (a) $\kappa=0$, the spanwise-uniform vortex shedding, and (b) $\kappa=4$ ($\beta h=0.49$), the peak near reattachment in figure~\ref{fig:coverage}(c). Dashed lines mark separation and reattachment. Each panel is scaled by its own maximum.}}
  \label{fig:spanshapes}
\end{figure}

The spanwise partition of the fluctuation energy (figure~\ref{fig:coverage}c) locates this oblique content in wavenumber. The spanwise modes are labelled by the index $\kappa$, with $\beta h=2\pi\kappa h/L_z$. The plane $\kappa=0$ carries $13.0\,\%$ of the fluctuation energy. The energy of the oblique modes, $\kappa\neq0$, is spread across the low wavenumbers, with a broad maximum at $\kappa=3$ in the whole window. The first five oblique modes hold $40\,\%$ of it. In the region $35<\xi<55$ around reattachment, the spectrum peaks at $\kappa=4$, that is $\beta h=0.49$ and $\lambda_{z}=12.9\,h=0.29\,L_{b}$. This wavelength is comparable to the roller spacing, which grows from $7$ to $11\,h$ between $\xi=32$ and $49$ in figure~\ref{fig:spanshapes}(a), so that $\lambda_{z}/\lambda_{x}=1.2$ at reattachment. This ratio lies at the low end of the range $1$ to $2$ reported for the spanwise deformation of rollers in separation bubbles \citep{Michelis2018,Kurelek2019}, and the deformation is visible in figure~\ref{fig:spanshapes}. The mode $\kappa=4$ is retained for the component-resolved analysis. The wall-parallel shapes of the wall-tangential velocity $\hat{u}_s$ of the two retained spanwise modes are given in figure~\ref{fig:spanshapes}. The mode $\kappa=0$ shows spanwise-coherent rollers and the mode $\kappa=4$ a spanwise-periodic pattern near reattachment.

\subsection{Production and transfer maps}\label{sec:maps}
\begin{figure}[!h]
  \centering\includegraphics[width=\linewidth]{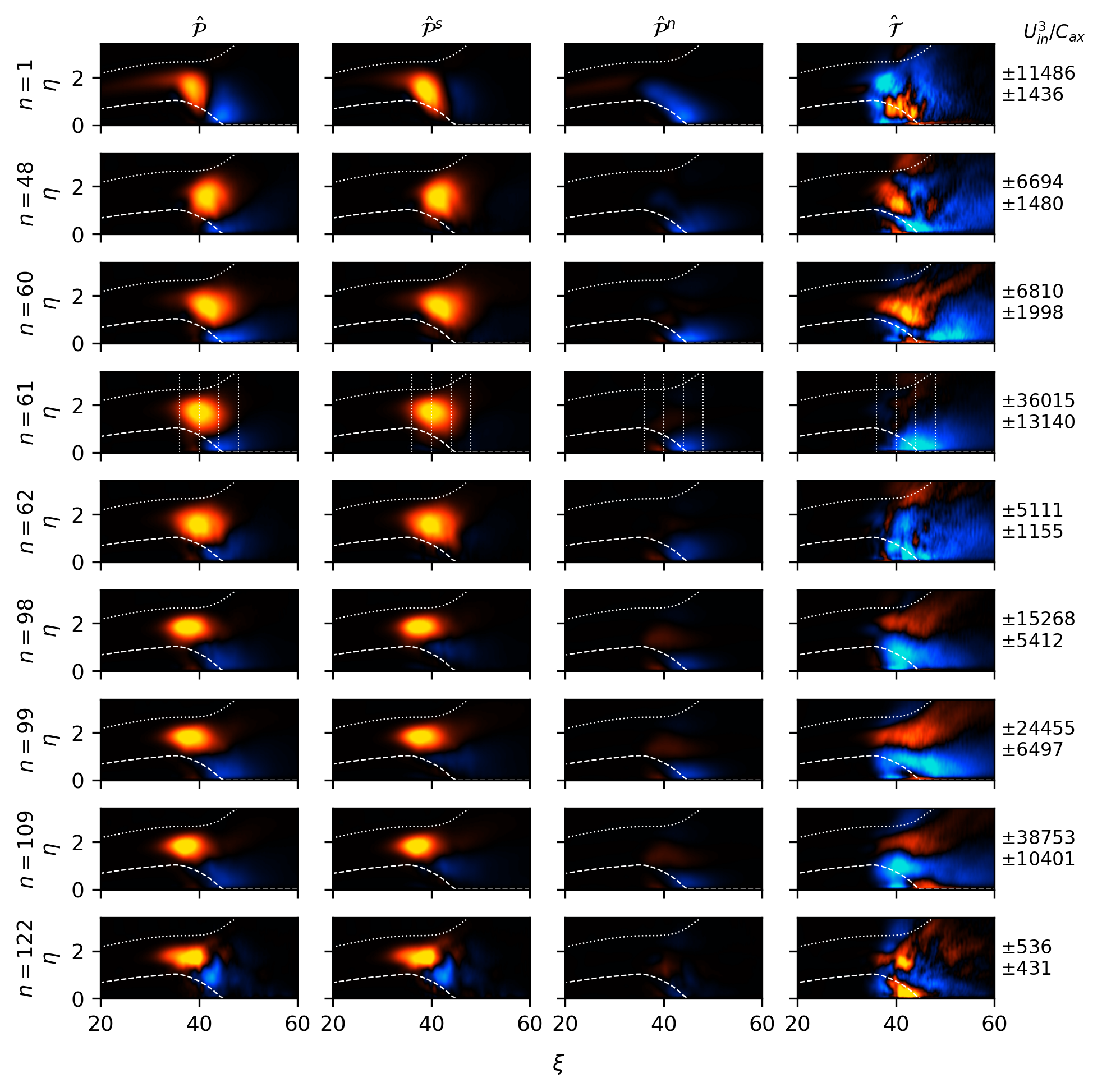}
  \caption{Production and transfer maps of the nine tonal modes at $\kappa=0$, one row per mode. Columns: production, shear production, normal production, and interscale transfer. The three production panels of a row share one colour scale. The transfer panel has its own. The two limits are printed at the right of each row in $U_{in}^{3}/C_{ax}$. Vertical dotted lines in the row $n=61$: the profile stations of figure~\ref{fig:budget61}.}
  \label{fig:maps}
\end{figure}

The integrated shares of table~\ref{tab:budgets} do not say where the energy enters and where it leaves. The maps of figure~\ref{fig:maps} locate the production, its two components, and the transfer of the nine modes. There is one row per mode, on the colour scales printed at its right. The production cell is one structure at every frequency. It is broad and centred at $\xi=38$ to $40$ for the modes up to $n=62$. It is compact, higher in the shear layer and $2$ to $3\,h$ upstream for the reattachment pair, the sum mode and the harmonic. Its amplitude differs by a factor of $70$ between the Kelvin-Helmholtz mode and the harmonic. The shear production reproduces the production map of every mode. The normal production is a region of extraction beneath the shear layer through reattachment, strongest for the flapping mode, whose production is a shear cell on the layer at $\xi=36$ to $42$ against a region of extraction at reattachment. The transfer maps separate the modes into two groups. The modes below the Kelvin-Helmholtz frequency gain at the transition front and lose downstream of it. The difference mode $n=48$ gains in the shear layer at $\xi=36$ to $46$, above the $\overline{u}=0$ line. The lower sideband $n=60$ gains in the same region, up to $\xi=48$, and in the outer part of the layer downstream of reattachment. The two modes lose beneath the shear layer from $\xi=42$ to $60$, below $\eta=1.5$, through and beyond reattachment. The flapping mode gains inside the reversed-flow region, beneath the $\overline{u}=0$ line at $\xi=38$ to $45$, and along the wall downstream of reattachment. The flapping mode loses in the shear layer, at $\xi=36$ to $42$ above $\eta=1.5$ and downstream of $\xi=42$ between $\eta=1$ and $2.5$. The Kelvin-Helmholtz mode and the modes above it, $n=61$ to $109$, lose in the shear layer from $\xi=38$ to $48$ and in the reversed-flow region beneath it. They gain above the shear layer downstream of reattachment, between $\eta=1.5$ and $3$ for $\xi=45$ to $60$, with smaller regions of gain near the wall around $\xi_{r}$. The upper sideband $n=62$ follows this group, with a weak gain above the layer. The harmonic $n=122$ has a fine-scale transfer of both signs. At the transition front, the low-frequency modes therefore gain, in the shear layer for the difference mode and the lower sideband and inside the reversed-flow region for the flapping mode, and the high-frequency modes lose. Above the shear layer downstream of reattachment, the high-frequency modes gain what the low-frequency modes lose. The transfer thus has a spatial structure that the frequency ordering of figure~\ref{fig:synthesis} does not show. Energy moves to the low frequencies inside the bubble and to the high frequencies above the reattached layer.

\begin{figure}[!h]
  \centering\includegraphics[width=\linewidth]{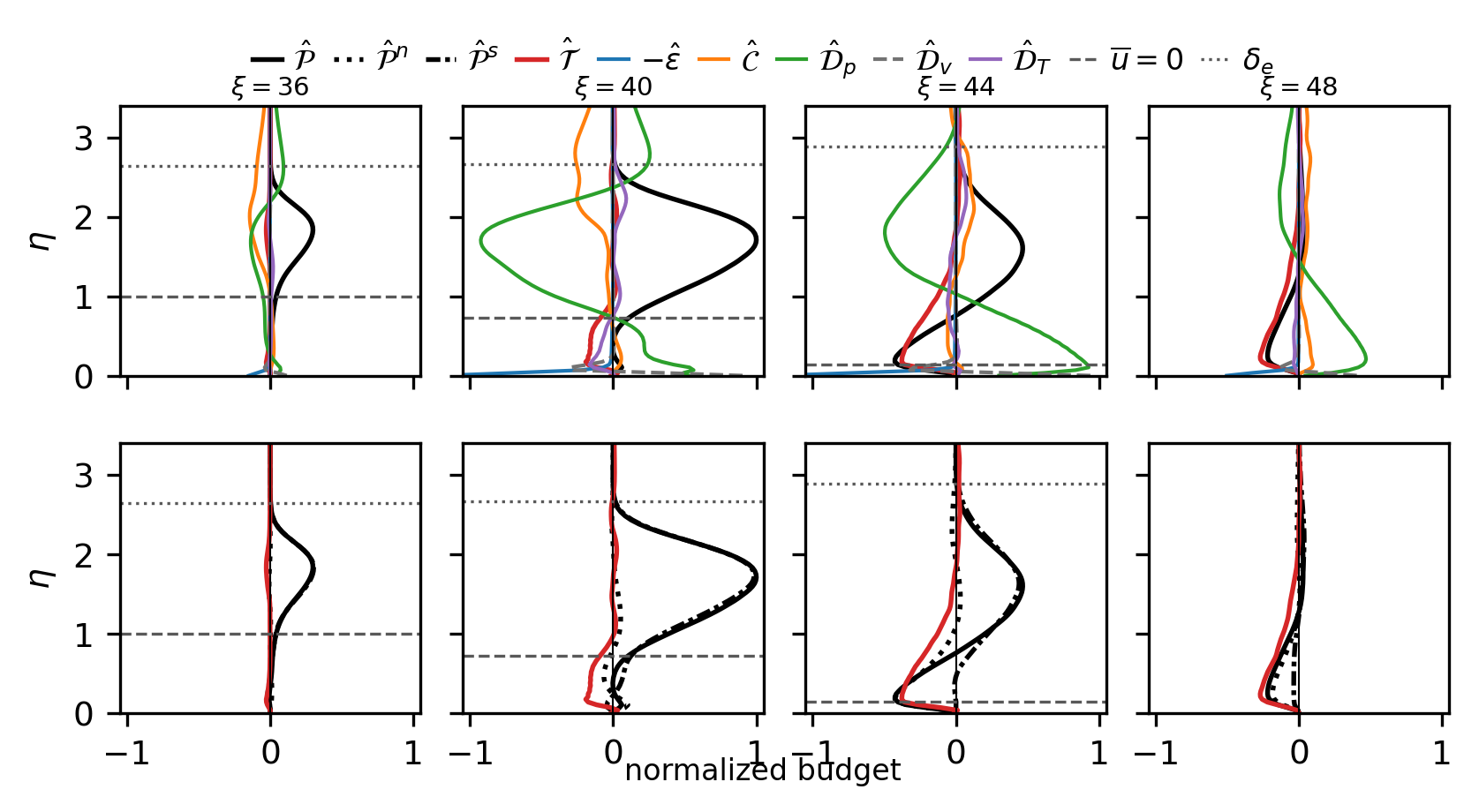}
  \caption{{Budget of the Kelvin-Helmholtz mode $n=61$ at $\kappa=0$ at the stations $\xi=36$, $40$, $44$, and $48$. Top row: the budget terms of (\ref{eq:spectral}). Bottom row: production, its two components of (\ref{eq:prodsplit}), and the transfer. All profiles are normalised by one value, the peak of the production map of figure~\ref{fig:maps}, $4.3\times10^{4}\,U_{in}^{3}/C_{ax}$. Dashed line: $\overline{u}=0$ height. Dotted line: $\delta_{e}$. The near-wall dissipation peak exceeds the axis range.}}
  \label{fig:budget61}
\end{figure}

The Kelvin-Helmholtz mode, the mode of largest production, serves to follow the balance from the shear layer to the wall. Its production is a compact region of positive $\hat{\mathcal{P}}$, called the production cell in the following, between the $\overline{u}=0$ height and the shear-layer edge $\delta_{e}$, centred at $\xi=40$ and $\eta=1.7$. The shear production carries the cell, with $\hat{\mathcal{P}}^{s}/\hat{\mathcal{P}}=1.28$ in the integral. The normal production is a weaker region of negative values beneath the shear layer, from $\xi=38$ through reattachment, with $\hat{\mathcal{P}}^{n}/\hat{\mathcal{P}}=-0.28$. This is the region where the mean flow decelerates (figure~\ref{fig:gradients}b) and the normal stresses return energy to the mean flow. The transfer is weak in the production cell and negative beneath the shear layer, strongest between $\xi=42$ and $50$ beneath $\eta=1$. The wall-normal profiles of all the terms of (\ref{eq:spectral}) at the four stations $\xi=36$, $40$, $44$ and $48$ marked on the row $n=61$ of figure~\ref{fig:maps}, normalised by the peak of its production map, are given in figure~\ref{fig:budget61}. They show how the production in the shear layer is balanced and which terms carry its effect toward the wall. At $\xi=40$ the production peaks in the shear layer and is balanced by the pressure transport, which removes energy from the shear layer and deposits it near the wall. At $\xi=44$ the pressure transport near the wall reaches $0.9$ of the production peak. It is balanced there by the transfer and the normal production, which have nearly identical profiles of $-0.4$ at $\eta=0.3$, and by the dissipation at the wall. At $\xi=48$ the same near-wall balance persists at half the amplitude. The shear production follows the total in the shear layer at every station. The normal production follows the transfer near the wall from $\xi=44$ onward. The pressure is therefore the term that brings the energy of the roll-up to the wall, where the transfer and the normal strains remove it. The transfer from the Kelvin-Helmholtz mode to the other modes takes place beneath the shear layer and not in the production cell.

\begin{figure}[!h]
  \centering\includegraphics[width=\linewidth]{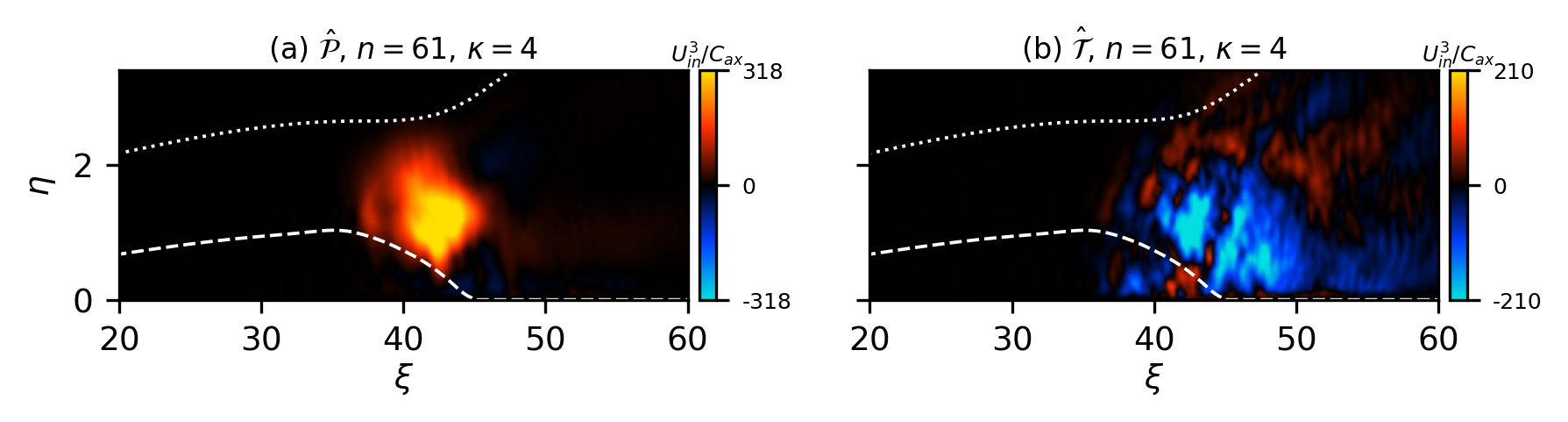}
  \caption{(a) Production and (b) transfer of the Kelvin-Helmholtz mode on the oblique plane $\kappa=4$, in $U_{in}^{3}/C_{ax}$. Dashed line: $\overline{u}=0$ height. Dotted line: $\delta_{e}$.}
  \label{fig:focus}
\end{figure}

The oblique plane is fed by the mean shear as well. At $\kappa=4$ the Kelvin-Helmholtz frequency has its own production cell at $\xi=38$ to $46$ (figure~\ref{fig:focus}a). Its transfer is fine-scaled through breakdown (figure~\ref{fig:focus}b). It is negative in the shear layer and positive downstream of reattachment, where the spanwise-periodic pattern of figure~\ref{fig:spanshapes}(b) is located. The oblique modes therefore receive energy from the mean flow directly, through their wall-tangential and wall-normal velocities. The transfer from the plane $\kappa=0$ is not their only source. How much each route supplies, and through which velocity component, is the subject of \S\,\ref{sec:spatial}.

\subsection{The routes of the tonal energy}\label{sec:transfer}
The transfers within the ranked triads of table~\ref{tab:triads} are called routes in what follows. The role of each mode in a triad follows from its efficiency $\gamma$ of (\ref{eq:eff}), given in table~\ref{tab:roles} for the four strongest triads. In the two sideband triads, $(61,1)\to62$ and $(60,1)\to61$, the Kelvin-Helmholtz mode is the donor, with $\gamma=-0.86$ and $-0.75$. The modes $62$ and $60$ are the recipients, with $\gamma=+0.50$ and $+0.95$. The flapping mode is the third member of both, the mode of the triad that is neither donor nor recipient. It loses in the lower-sideband triad, $\gamma=-0.19$, and gains in the upper one, $\gamma=+0.41$. It is therefore not a catalyst in the strict sense of \citet{Baj2017}. In the triad $(1,98)\to99$ the flapping mode has $\gamma=-0.02$ and is a catalyst in that sense. The mode $98$ donates to the mode $99$ with $\gamma=-0.55$ and $+0.57$. In the triad $(48,61)\to109$, the sum mode $109$ donates with $\gamma=-0.23$, the difference mode $48$ receives with $\gamma=+0.37$. The Kelvin-Helmholtz mode is the catalyst with $\gamma=-0.12$. This triad transfers energy from a higher to a lower frequency and is referred to as the backward triad. The harmonic triad $(61,61)\to122$, marked H in figure~\ref{fig:selection}(a), has an exchange comparable to the self-interaction entries of individual broadband bins and does not enter the ranking of table~\ref{tab:triads}. For the present case, the tonal energy moves by sideband and backward transfers and not by harmonic generation. A sideband transfer with a low-frequency third member is also the route of vortex pairing in transitional jets \citep{Nekkanti2025b} and of the low-frequency content of a transitional shock-induced bubble \citep{BenHassanSaidi2025}. The backward triad transfers energy between two modes of the donor band of figure~\ref{fig:synthesis}.

\begin{table}
  \centering
  \caption{Efficiency $\gamma$ of (\ref{eq:eff}) of the three modes of the four strongest triads of table~\ref{tab:triads}, summed over the spanwise modes. The donor has $\gamma<0$, the recipient $\gamma>0$, and a catalyst $|\gamma|\ll1$.}
  \label{tab:roles}
  \begin{tabular}{lccc}
  \toprule
  triad (donor $\to$ recipient) & donor & catalyst & recipient \\
  \midrule
  $61\to60$, lower sideband & $61$: $-0.75$ & $1$: $-0.19$ & $60$: $+0.95$ \\
  $61\to62$, upper sideband & $61$: $-0.86$ & $1$: $+0.41$ & $62$: $+0.50$ \\
  $98\to99$, reattachment pair & $98$: $-0.55$ & $1$: $-0.02$ & $99$: $+0.57$ \\
  $109\to48$, backward & $109$: $-0.23$ & $61$: $-0.12$ & $48$: $+0.37$ \\
  \bottomrule
  \end{tabular}
\end{table}

Where the energy exported by the four tonal donors goes is given in table~\ref{tab:destination}. For each donor, the negative entries of the corresponding row of $A$ are summed separately over three groups of recipient modes: the tonal set, the non-tonal modes at lower frequencies, and the non-tonal modes at higher frequencies. Of the energy exported by the Kelvin-Helmholtz mode, $18.9\,\%$ goes to the tonal set and the remainder splits into $55.4\,\%$ toward higher frequencies and $44.6\,\%$ toward lower frequencies. The sum mode exports $66.4\,\%$ of its energy to lower frequencies, the difference mode $84.1\,\%$ to higher frequencies, and the flapping mode all of it to higher frequencies. Transfer to lower frequencies has been reported in the spot-growth region of bypass transition \citep{Yao2022} and in the near wake of a cylinder \citep{Yeung2026}.

\begin{table}
  \centering
  \caption{Destination of the energy exported by the tonal donors, summed over the spanwise modes, as percentages of the energy the donor gives to all other modes: to the modes of the tonal set, and to modes of lower and of higher frequency than the donor. The last two columns sum to $100\,\%$, and the first is a subset of both.}
  \label{tab:destination}
  \begin{tabular}{lccc}
  \toprule
  donor & tonal set & lower frequencies & higher frequencies \\
  \midrule
  $n=61$, Kelvin-Helmholtz & $18.9$ & $44.6$ & $55.4$ \\
  $n=109$, sum mode & $17.3$ & $66.4$ & $33.6$ \\
  $n=48$, difference mode & $4.9$ & $15.9$ & $84.1$ \\
  $n=1$, flapping & $17.5$ & $0.0$ & $100.0$ \\
  \bottomrule
  \end{tabular}
\end{table}

\begin{figure}[!h]
  \centering\includegraphics[width=\linewidth]{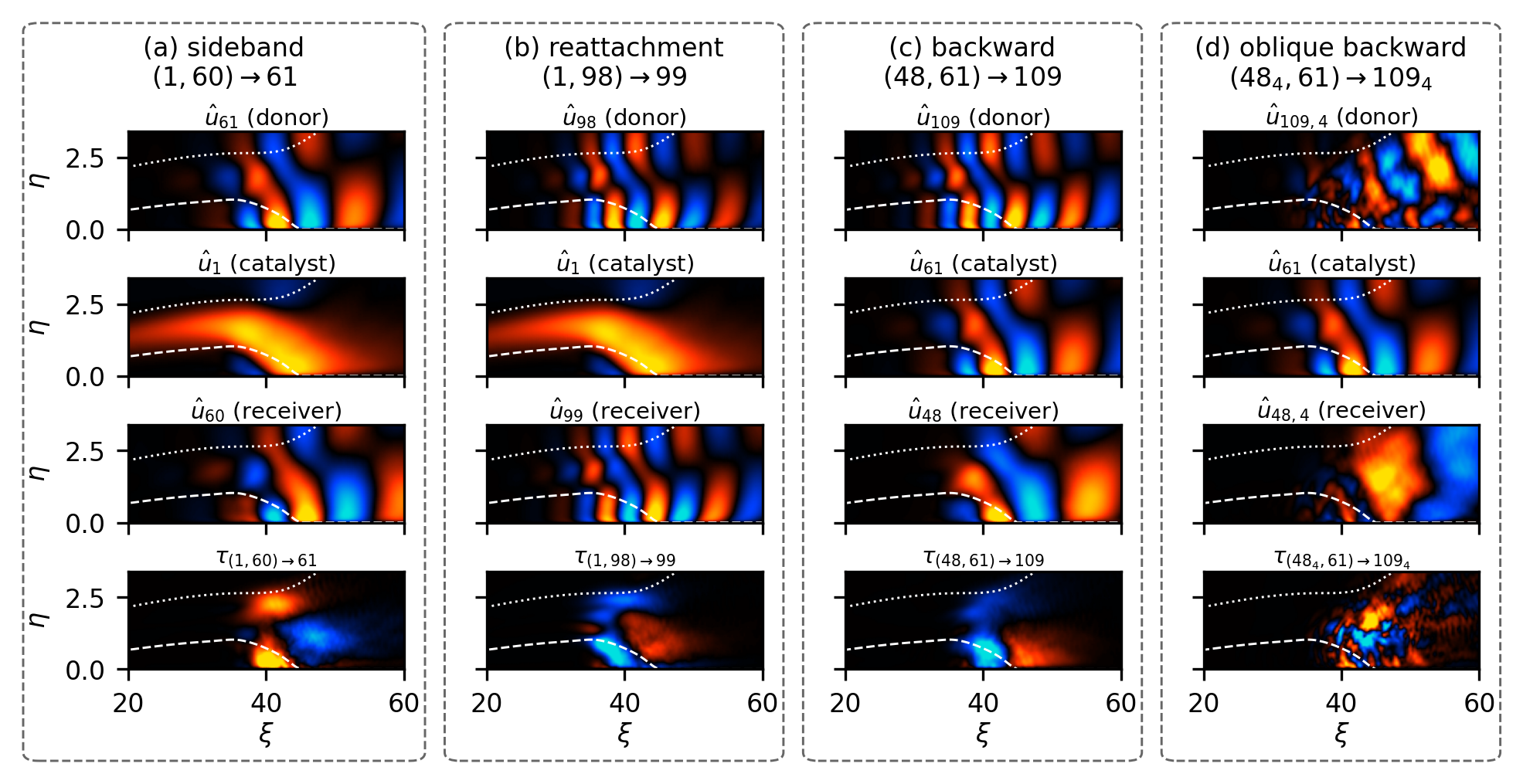}
  \caption{{Anatomy of four routes, one column per triad, named in the box above each column, with all modes at $\kappa=0$ except in column (d), where donor and recipient are at $\kappa=4$. Rows one to three: shapes of the wall-tangential velocity $\Re[\hat{u}_{s}]$ of the donor, the catalyst, and the receiver. Bottom row: transfer density $\tau$ of (\ref{eq:tau}), positive where the receiver gains. Shapes are phase-aligned to their peak. Each panel is scaled by its own maximum.}}
  \label{fig:routes}
\end{figure}

The efficiencies of table~\ref{tab:roles} are integrals. They do not show where in the bubble a route operates, nor whether the shape of the third member accounts for its role. Both are read from the shapes of the three modes and the density of their exchange. They are given in figure~\ref{fig:routes} for four routes, one column per triad: the lower-sideband, reattachment and backward triads of table~\ref{tab:roles} and the oblique backward triad of figure~\ref{fig:selection}(b). The three upper rows show the wall-tangential velocity $\Re[\hat{u}_s]$ of the three modes of the triad, in the order donor, catalyst, and recipient. The bottom row shows the transfer density $\tau$ of (\ref{eq:tau}), positive where the recipient gains. The flapping mode, catalyst in columns (a) and (b), is a smooth structure that follows the shear layer along the whole bubble. The Kelvin-Helmholtz mode, donor in column (a) and catalyst in columns (c) and (d), is a wave train that grows from $\xi=30$ and reaches the wall downstream of $\xi=42$. In columns (a) and (b), the donor and the recipient are wave trains of nearly the same wavelength, offset by the frequency of the catalyst. The recipient of column (c), the mode $n=48$, is a structure of longer wavelength that fills the reattached region, $\xi>45$. Its oblique counterpart in column (d), the mode $n=48$ at $\kappa=4$, is the spanwise-periodic pattern of figure~\ref{fig:spanshapes}(b). The transfer densities are located at the transition front. In column (a), the density is positive in the shear layer from $\xi=38$ to $42$ and negative from $\xi=43$ to $52$. In column (b), it is negative along the shear layer from $\xi=36$ to $44$ and positive at and beneath reattachment. In column (c), it is negative at the front and positive near the wall downstream of $\xi_{r}$, signed for the sum mode, whose loss is the gain of the mode $48$. In column (d), it is a fine-grained exchange through breakdown, between $\xi=38$ and $52$, in which the two-dimensional Kelvin-Helmholtz mode is the catalyst of the transfer between the two oblique modes.

\subsection{Velocity channels of the routes}\label{sec:spatial}
The routes of \S\,\ref{sec:transfer} state which modes exchange energy and where. They do not state through which velocity component the exchange takes place, which is the information needed to relate the transfer to the three-dimensionalisation of the rollers, that is, the growth of spanwise velocity in the oblique modes. The channels of (\ref{eq:tau9}) give this information. The spanwise velocity is the component to follow. Indeed, with $U_w=0$ and no spanwise dependence of the mean, the production of $|\hat{u}_w|^{2}$ in (\ref{eq:prodsplit}) vanishes for every mode. The spanwise velocity of a mode can gain energy from two terms only. One is the pressure-strain redistribution within the mode. It cancels in the component sum of (\ref{eq:spectral}) and is not evaluated here. The other is the nonlinear transfer, whose component-resolved part is the row $i=w$ of (\ref{eq:tau9}). This is the modal counterpart of the budget of the spanwise stress in wall turbulence. That stress has no production and is fed by pressure-strain and transport \citep{Mansour1988}. The growth of spanwise velocity in the rollers is a marker of the onset of their breakdown. Linear theory describes it in two ways. The first is the amplification of oblique waves by the mean flow \citep{Craik1971,Michelis2018,Rodriguez2021}. The second is the secondary instability of the periodic rollers, elliptic in the cores and hyperbolic in the braids \citep{Jones2008,Marxen2013}. In the modal budget, the first appears as the production of the oblique modes, which feeds their wall-tangential and wall-normal velocities only, since the production of their spanwise velocity vanishes. The second appears as a triad in which the two-dimensional roller is the catalyst, because a disturbance that grows on the periodic base state formed by the roller exchanges energy with the other modes through the velocity of the roller. The budget measures both.

The linear route is weak on the plane $\kappa=4$. The production of the oblique difference mode $48_4$ is $11\,\%$ of that of its two-dimensional twin. The production of the oblique sum mode $109_4$ is $0.3\,\%$ of that of the mode $109$. The difference mode exports $80\,\%$ of its production through the transfer, as the two-dimensional modes do. The sum mode has a net transfer of $2\,\%$ of its production. Its exchanges with the other modes nearly cancel in the integral. Which of them carry its spanwise velocity is visible only in the channels.

The channel $\hat{\mathcal{T}}_{sn}$ is the wall-tangential velocity advected by the wall-normal velocity. Summed over all triads of a mode, it carries $44\,\%$ of the transfer of the Kelvin-Helmholtz mode, $90\,\%$ of the sum mode, $71\,\%$ of the difference mode, and $84\,\%$ of the flapping mode. The transfer thus results mostly from the joint action of the wall-normal velocity of the catalyst and the wall-normal gradient of the wall-tangential velocity of the donor, $\partial\hat{u}^{\,l}_{s}/\partial n$ in (\ref{eq:tau9}). The spanwise-advection channel $\hat{\mathcal{T}}_{sw}$ is non-zero only through oblique catalysts, because the catalyst supplies the advecting velocity and a two-dimensional catalyst has none along the span. It carries $41\,\%$ of the transfer of the Kelvin-Helmholtz mode. Two-fifths of the energy that the two-dimensional rollers export thus leave them by spanwise advection, through a velocity that the plane $\kappa=0$ does not contain. The spanwise redistribution is not confined to the oblique modes. It is the second route out of the roller, after the wall-normal advection. The exchange of the tonal modes with the oblique modes is therefore carried by the spanwise velocity. The channels of the oblique backward triad are examined in detail below for that reason.

\begin{figure}[!h]
  \centering\includegraphics[width=\linewidth]{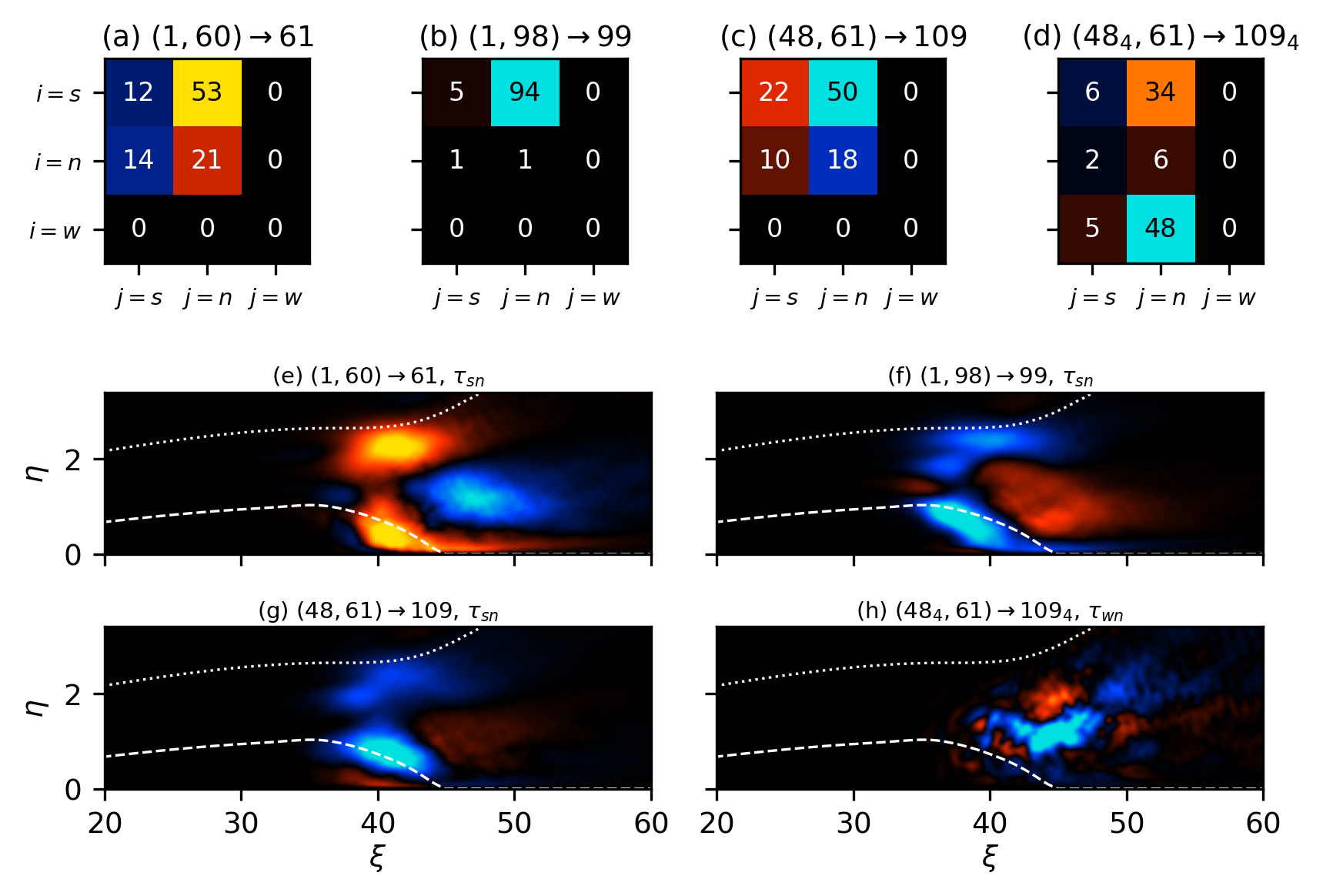}
  \caption{{Velocity channels of the routes. (a to d) The nine channel integrals $\int\tau_{ij}\,\mathrm{d}V$ of (\ref{eq:tau9}) for the lower-sideband, reattachment, and backward triads at $\kappa=0$, and the oblique backward triad with donor and recipient at $\kappa=4$. Rows: recipient component $i$. Columns: advecting component $j$ of the catalyst. Numbers: share of each channel in the summed magnitude, in per cent, rounded to the unit. Tiles are scaled by their own maximum. (e to h) Density of the dominant channel of each route, named in the title with the notation of (\ref{eq:tau9}), scaled by its own maximum, positive where the recipient gains.}}
  \label{fig:channels}
\end{figure}

Route by route, one channel dominates. This follows from the decomposition of the four routes of figure~\ref{fig:routes} into the nine channels of (\ref{eq:tau9}) in figure~\ref{fig:channels}. The tiles of panels (a) to (d) are the integrals $\int_V\tau_{ij}\,\mathrm{d}V$ of the nine channel densities of one triad. The row index $i$ is the velocity component whose energy is exchanged, carried by the donor through its gradient $\partial\hat{u}^{\,l}_{i}/\partial x_j$ in (\ref{eq:tau9}) and received by the recipient $\hat{u}^{\,n}_{i}$. The column index $j$ is the component of the catalyst velocity $\hat{u}^{\,m}_{j}$ that advects it, along $x_j$. The number in each tile is the share of that channel in $\sum_{ij}|\int_V\tau_{ij}\,\mathrm{d}V|$, in per cent. The maps of panels (e) to (h) are the densities of the largest channel of each route. The column advected by the spanwise velocity of the catalyst vanishes in all four routes, because the catalyst is two-dimensional in each of them. For the two-dimensional triads, panels (a) to (c), the spanwise row vanishes as well, so the exchange is contained in the four tiles of the tangential and normal components. The channel $\tau_{sn}$, the wall-tangential velocity of the donor advected by the wall-normal velocity of the catalyst, carries $53\,\%$ of the sideband exchange, $94\,\%$ of the reattachment exchange, and $50\,\%$ of the backward exchange. In the oblique backward triad, panel (d), the donor and the recipient are oblique. Its channel $\tau_{wn}$, the spanwise velocity of the donor advected by the wall-normal velocity of the catalyst and received by the recipient, carries $48\,\%$ of the exchange of the oblique triad. The three channels of the spanwise-recipient row together carry $53\,\%$ of it. Panels (e) to (h) compare the map of the largest channel with the map of the total density $\tau$ of figure~\ref{fig:routes}. The two maps are similar for all four routes. The correlation coefficient between the dominant channel and the total density in the window is $0.87$, $0.98$, and $0.78$ for the sideband, reattachment, and backward routes, whose dominant channel is $\tau_{sn}$. It is $0.63$ for the oblique route, whose dominant channel is $\tau_{wn}$. The advecting component is the wall-normal velocity in every route. In a turbulent channel, the interaction that bends the near-wall streaks, a spanwise velocity advecting the streamwise velocity of the streak, is advected by the spanwise velocity \citep{ChenGarciaMayoral2026}. The difference with the present routes is set by the catalysts. A roller and the flapping of a shear layer advect the other modes with their wall-normal velocity. A meandering streak advects them with its spanwise velocity. The channel of a route is thus fixed by the kinematics of its catalyst and not by the recipient.

\begin{figure}[!h]
  \centering\includegraphics[width=\linewidth]{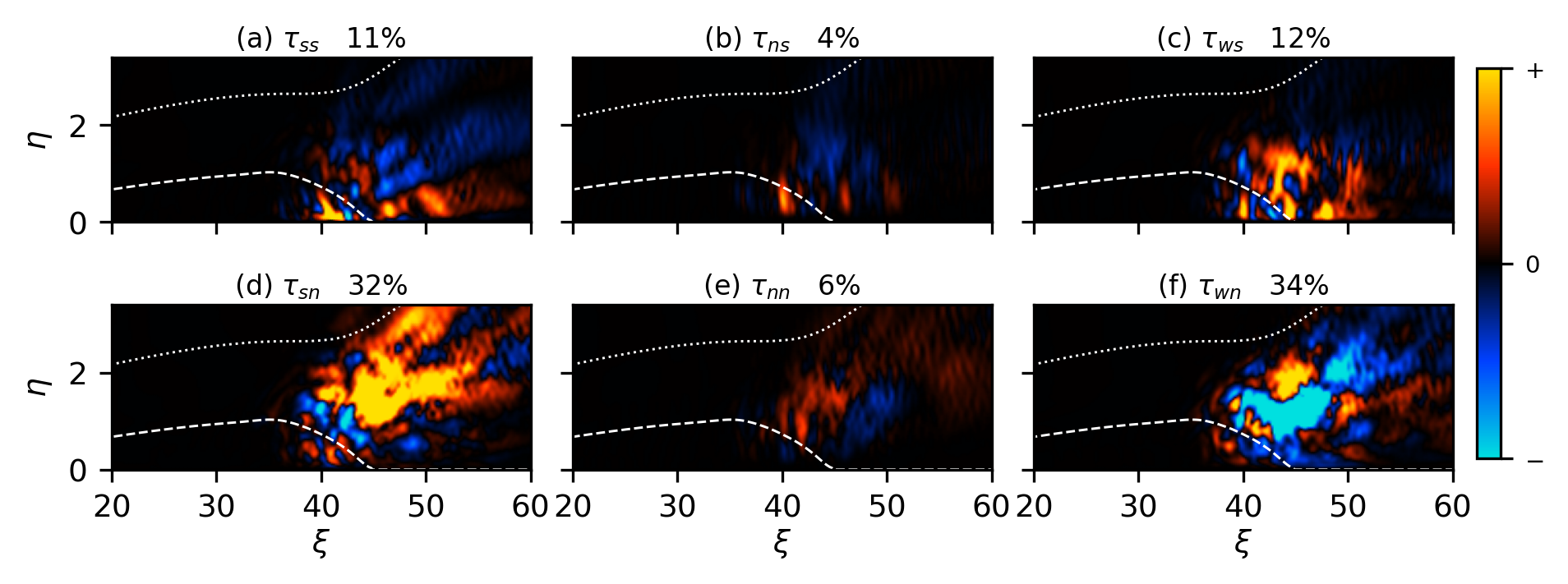}
  \caption{{The six non-vanishing channel densities $\tau_{ij}$ of the oblique backward triad $(48_4,61)\to109_4$ on one colour scale, positive where the sum mode $109_4$ gains. Columns: recipient component. Rows: advecting component of the catalyst, wall-tangential in (a) to (c) and wall-normal in (d) to (f). Titles give the channel in the notation of (\ref{eq:tau9}) and its share in the pointwise absolute density in the window.}}
  \label{fig:grid9}
\end{figure}

Within the oblique triad, the exchange of spanwise velocity is confined to one band. This is seen in its six non-vanishing channel densities, drawn on one colour scale in figure~\ref{fig:grid9} with the recipient component along the columns and the advecting component along the rows. Panels (a) to (c) are the channels advected by the wall-tangential velocity of the catalyst and panels (d) to (f) those advected by its wall-normal velocity. Panels (d) and (f) occupy the same band, inclined through breakdown from $\xi=38$ near the wall to $\xi=50$ at $\eta=2$, with opposite signs. In this band, the oblique sum mode gains wall-tangential energy, panel (d), and loses more spanwise energy, panel (f), so the exchange of the oblique triad is carried by the spanwise velocity of the recipient. The channels advected by the wall-tangential velocity, panels (a) to (c), change sign along the band, so they hold $12\,\%$ of the integral of $|\tau_{ij}|$ in the window, summed over the six channels, but only $5\,\%$ of the integral of $\tau_{ij}$.

\begin{figure}[!h]
  \centering\includegraphics[width=0.62\linewidth]{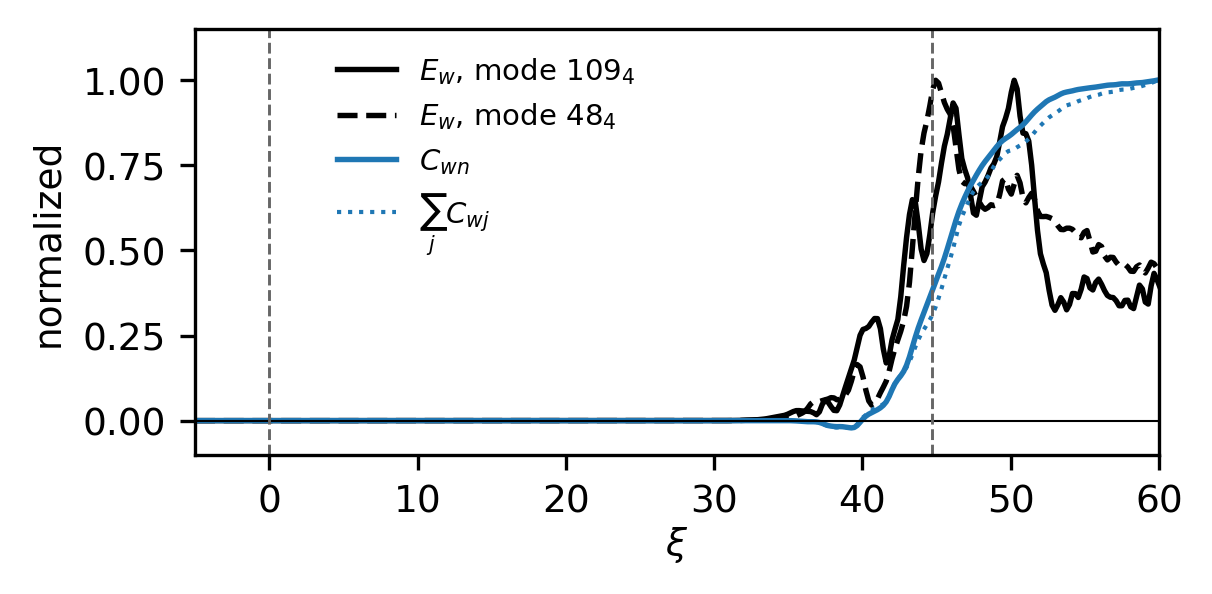}
  \caption{Spanwise-velocity energy $E_w$ of (\ref{eq:ew}) of the oblique sum mode $109_4$ and difference mode $48_4$, and accumulated exchange $C_{wn}$ of (\ref{eq:cum}) of the sum mode through the channel $\tau_{wn}$, together with its sum $\sum_j C_{wj}$ over the three advecting components. All curves are normalised by their maximum. Vertical lines: separation and reattachment.}
  \label{fig:mech}
\end{figure}

The band of figure~\ref{fig:grid9} is where the oblique modes acquire spanwise velocity. This is established by comparing the growth of the spanwise-velocity energy of the two oblique modes with the streamwise accumulation of the exchange through $\tau_{wn}$ (figure~\ref{fig:mech}). The spanwise-velocity energy of a mode at a station is
\begin{equation}
  E_w(\xi)=\int_0^{\delta_e}|\hat{u}_w|^{2}\,\mathrm{d}\eta,
  \label{eq:ew}
\end{equation}
and the accumulated exchange through a channel from $\xi_0=20$ is
\begin{equation}
  C_{ij}(\xi)=-\int_{\xi_0}^{\xi}\int_0^{\delta_e}\tau_{ij}\,\mathrm{d}\eta\,\mathrm{d}\xi',
  \label{eq:cum}
\end{equation}
signed as a loss of the sum mode. The figure shows $E_w$ of the sum and difference modes, $C_{wn}$, and the sum of $C_{wj}$ over the three advecting components, each normalised by its maximum. The two accumulated curves coincide, so the spanwise exchange of the sum mode occurs through $\tau_{wn}$. The spanwise energy of both modes grows between $\xi=38$ and $46$ and decays downstream of $\xi=50$. The exchange accumulates between $\xi=40$ and $52$, in the same interval in which the mean shear produces the oblique modes. The two routes are compared by their integrals. The entry of the oblique triad is $-0.55\times10^{-8}\,U_{in}^{3}C_{ax}^{2}$. It is $5.4\,\%$ of the spanwise-uniform triad of the same frequencies, $-1.03\times10^{-7}\,U_{in}^{3}C_{ax}^{2}$. That triad carries $63\,\%$ of the entry of table~\ref{tab:triads}. The oblique entry is $0.37$ of the production of the oblique sum mode. One triad with the two-dimensional Kelvin-Helmholtz mode as catalyst therefore moves out of the oblique sum mode and into the oblique difference mode an amount of spanwise energy comparable to the energy the mean shear gives the sum mode. Since the net transfer of the sum mode is $2\,\%$ of its production, other triads return this amount. The spanwise velocity of the oblique pair is thus neither produced by the mean flow nor given by the roller, which supplies only the advecting velocity in every channel of figure~\ref{fig:grid9}. The spanwise velocity is exchanged between the oblique modes through the wall-normal velocity of the roller. The linear routes supply the oblique modes with wall-tangential and wall-normal velocity, at $11\,\%$ of the two-dimensional level for the difference mode. The triads with the roller as catalyst move the spanwise velocity between them. The source of the spanwise energy of the pair as a whole, pressure-strain within each mode or transfer from modes outside the tonal set, is not separated by the present budget.

\begin{figure}[!h]
  \centering\includegraphics[width=\linewidth]{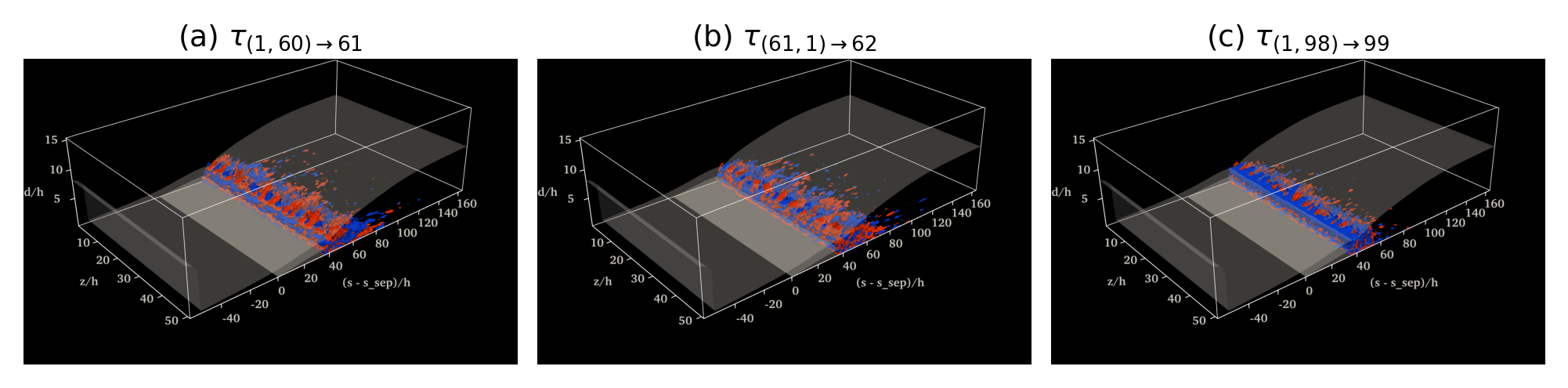}
  \caption{{Transfer densities in three dimensions of the three routes with the flapping mode as the third member, containing all spanwise triads of the three frequencies. (a) Lower-sideband, (b) upper-sideband, and (c) reattachment triad at the same iso-surface level, red where the recipient gains and blue where it loses. Light grey: the mean reversed-flow region, bounded by the $\overline{u}=0$ surface. Dark grey: $\delta_{e}$.}}
  \label{fig:tau3d}
\end{figure}

The two planes $\kappa=0$ and $4$ do not show how the exchange is organised along the span. The densities of the frequency triads evaluated in $(x,y,z)$ as described in \S\,\ref{sec:tensor} do, because each of them contains every spanwise triad $(\kappa_l,\kappa_m)\to\kappa_l+\kappa_m$ of the three frequencies. Their iso-surfaces for the three routes with the flapping mode as third member are drawn at one iso level in figure~\ref{fig:tau3d}, red where the recipient gains and blue where it loses. The three densities are located in a thin sheet at the transition front, $\xi=36$ to $50$, around the breakdown of the rollers. The two sideband exchanges, panels (a) and (b), are quasi-periodic along the span at the wavelength of the mode $\kappa=4$ of figure~\ref{fig:spanshapes}(b). The reattachment exchange, panel (c), is negative along a spanwise-coherent sheet in the shear layer and positive in scattered patches downstream of it. The flapping mode, spanwise uniform at $\kappa=0$, thus organises an exchange that is modulated along the span at the roller wavelength as soon as its oblique components are included. This is the three-dimensional form of the spanwise-modulated flapping measured by \citet{Steinfurth2025}.

\begin{figure}[!h]
  \centering\includegraphics[width=\linewidth]{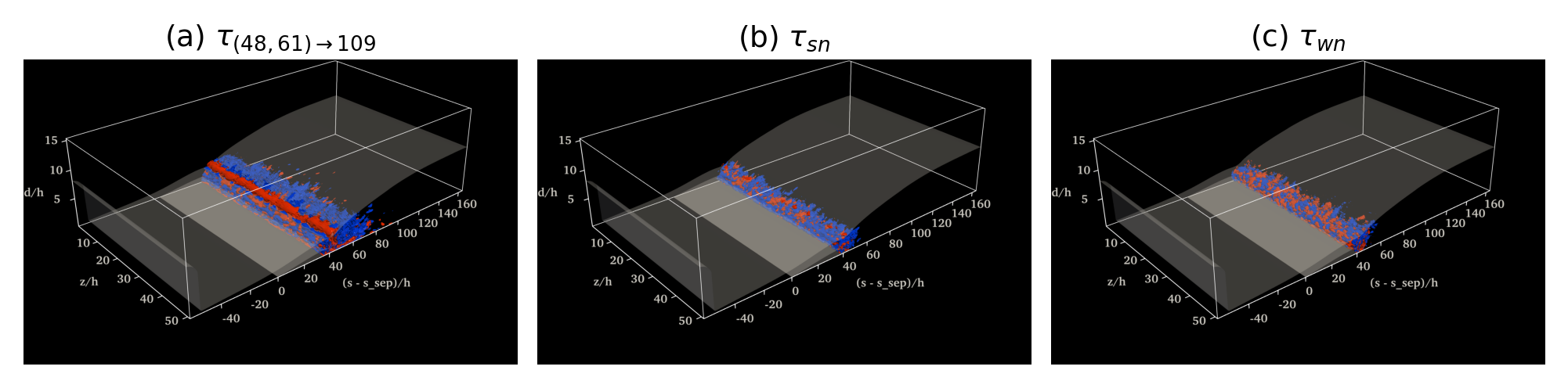}
  \caption{{Transfer density in three dimensions of the backward triad $(48,61)\to109$, containing all spanwise triads of the three frequencies. (a) Total density $\tau$, (b) channel $\tau_{sn}$, and (c) channel $\tau_{wn}$, each at its own iso level, red where the recipient gains and blue where it loses. Light grey: the mean reversed-flow region, bounded by the $\overline{u}=0$ surface. Dark grey: $\delta_{e}$.}}
  \label{fig:tau3dbw}
\end{figure}

The backward triad, with the Kelvin-Helmholtz mode as catalyst, is organised differently. Its density, figure~\ref{fig:tau3dbw}(a), is negative along spanwise-coherent sheets at the front and positive along a spanwise-coherent band near the wall downstream of reattachment. Its two channels are given at their own iso levels. The channel $\tau_{sn}$, panel (b), reproduces the band and the sheets of panel (a). The channel $\tau_{wn}$, panel (c), an order of magnitude weaker, occupies the same front with finer spanwise structure. The spanwise organisation of a route is therefore set by its catalyst. With the flapping mode as third member, the exchange is periodic along the span. With the roller as catalyst, it is spanwise coherent where it is carried by $\tau_{sn}$. It is spanwise fine where it is carried by $\tau_{wn}$, the channel of the oblique modes.

\begin{figure}[!h]
  \centering\includegraphics[width=\linewidth]{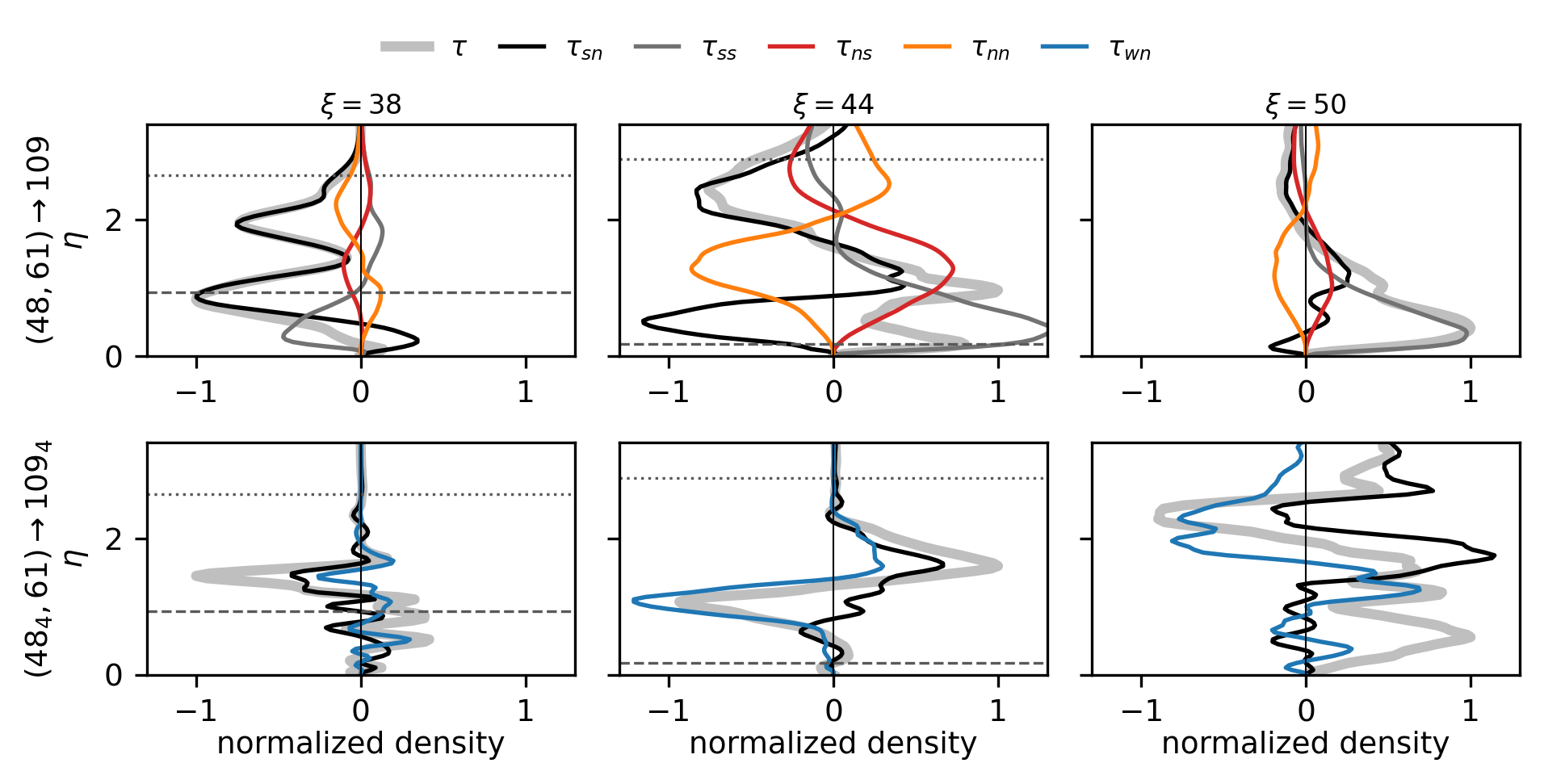}
  \caption{{Wall-normal profiles of the channel densities $\tau_{ij}$ of (\ref{eq:tau9}) for the backward triad (top) and its oblique twin (bottom) at $\xi=38$, $44$ and $50$. Grey: total density $\tau$. Coloured: the non-vanishing channels of the backward triad, and the channels $\tau_{sn}$ and $\tau_{wn}$ of the oblique triad. Each panel is normalised by the peak of the total. Dashed line: $\overline{u}=0$ height. Dotted line: $\delta_{e}$.}}
  \label{fig:chanprofiles}
\end{figure}

The wall-normal structure of the two backward triads completes the picture. The profiles of their channel densities are given in figure~\ref{fig:chanprofiles}, top row the two-dimensional triad and bottom row its oblique counterpart, at the three stations $\xi=38$, $44$ and $50$, normalised by the peak of the total density $\tau$ at each station. In the two-dimensional triad, four channels are non-zero. At $\xi=38$ the channel $\tau_{sn}$ follows the total. At $\xi=44$ it follows the total in the shear layer, above $\eta=1$, whereas near the wall the positive total is carried by $\tau_{ss}$ and, above $\eta=0.5$, by $\tau_{ns}$, while $\tau_{sn}$ and $\tau_{nn}$ are negative. At $\xi=50$ the total is carried by $\tau_{ss}$, with $\tau_{sn}$ small and negative. The two-dimensional exchange thus changes channel along the front. It is advected by the wall-normal velocity of the roller in the shear layer at the start of breakdown and by its wall-tangential velocity near the wall at reattachment, where the roller has reached the wall (figure~\ref{fig:modes}). For the oblique triad, only the two channels that carry its exchange, $\tau_{sn}$ and $\tau_{wn}$, are drawn. At $\xi=38$ the exchange is still carried by $\tau_{sn}$. At $\xi=44$ and $50$ the channel $\tau_{wn}$ is the largest and has the opposite sign of $\tau_{sn}$, which is the pattern of figure~\ref{fig:grid9}(d) and (f) read along the wall-normal direction. The exchange that makes the rollers three-dimensional thus occurs through one channel: the spanwise velocity of the oblique modes advected by the wall-normal velocity of the two-dimensional Kelvin-Helmholtz mode, between $\xi=40$ and $52$. It starts in the shear layer, at the height of the roller cores, and not at the wall. This is the location of the elliptic instability of the cores \citep{Marxen2013} and not that of the near-wall streaks of a geometry-induced separation \citep{Fang2024}.

Three points of this section bear on how the breakdown of the rollers is described. The first concerns the energy source of the oblique modes. A secondary-instability calculation on the frozen roller \citep{Marxen2013,Rodriguez2021} gives the growth rate of the oblique disturbance and attributes its growth to the periodic base state. The budget separates the two things the base state can do. It can give energy, as a donor, or it can advect, as a catalyst. In the backward triad, the roller has an efficiency of $-0.12$ and is a catalyst. In the oblique triad, it supplies the advecting velocity of every channel. The energy of the oblique modes has two sources. The mean shear produces them, at $11\,\%$ of the two-dimensional level for the difference mode. The rest comes from exchanges among the oblique modes themselves. The roller organises these exchanges without feeding them. The second point concerns the velocity channel. The exchange that gives the oblique modes their spanwise velocity goes through $\tau_{wn}$ in one band at the transition front. The two-dimensional exchanges go through $\tau_{sn}$. The advecting velocity is the wall-normal one in both, because the catalysts are rollers and the flapping of the shear layer. In wall turbulence, the interaction that bends the streaks is advected by the spanwise velocity, because the catalysts are the streaks themselves \citep{ChenGarciaMayoral2026}. The channel decomposition thus distinguishes the two classes of breakdown by their catalyst, before any structure is identified in the fields. The third point concerns the sparsity of the exchange. Half of it is carried by $7\,\%$ of the donor-catalyst pairs and $90\,\%$ by $46\,\%$. \citet{ChenGarciaMayoral2026} sustain channel turbulence with $30\,\%$ of the interactions and find deviations near the wall at $15\,\%$. The bubble lies in the same range. Its linear response is low-rank \citep{Yeh2020}, but the nonlinear forcing of its tonal modes is not. Moreover, $41\,\%$ of the transfer of the roller leaves it through oblique catalysts that a two-dimensional description does not contain.

\section{Conclusions}\label{sec:conclusions}
The kinetic energy budget of a laminar separation bubble on a compressor blade has been resolved mode by mode in frequency and spanwise wavenumber. The nonlinear transfer between modes has been evaluated as a triadic tensor that assigns a signed exchange and a role, donor, recipient, or catalyst, to each mode of every admissible triad. Each exchange has been split into nine velocity channels and located in the bubble by its density.

Summed over all modes, every frequency below $2.2$ times the Kelvin-Helmholtz frequency is a net donor and every frequency above is a net receiver. Nine tonal modes were identified from the strongest triads. They hold $8.2\,\%$ of the fluctuation energy and take part in $18.6\,\%$ of the energy exchanged, but $90\,\%$ of the exchange requires $389$ frequency modes and $46\,\%$ of the donor-catalyst pairs. All tonal modes are fed by the mean shear in one production cell on the separated shear layer. The cell moves upstream as the frequency increases. All export $69$ to $82\,\%$ of their production through the transfer at the transition front. The normal strains return $8$ to $35\,\%$ of the shear production to the decelerating mean. For the flapping mode, they return $63\,\%$, because the flapping is a roller-like structure with $|\hat{u}_n|>|\hat{u}_s|$ and not a streamwise-elongated one.

The Kelvin-Helmholtz mode exports $44.6\,\%$ of its energy to lower frequencies, the strongest of these transfers being the backward triad $(48,61)\to109$, in which it acts as a catalyst. The sidebands are fed through the flapping mode, which thereby broadens the spectrum around the Kelvin-Helmholtz peak. Harmonic generation is negligible. In every two-dimensional triad, one velocity channel, the wall-tangential velocity of the donor advected by the wall-normal velocity of the catalyst, carries $50$ to $94\,\%$ of the exchange with a density similar to that of the whole triad. The spanwise-advection channel, which only oblique catalysts can supply, carries $41\,\%$ of the transfer of the Kelvin-Helmholtz mode. On the oblique plane $\kappa=4$, the mean shear produces the oblique difference mode at $11\,\%$ of the two-dimensional level and the oblique sum mode at $0.3\,\%$. The wall-normal velocity of the Kelvin-Helmholtz mode moves spanwise energy from the oblique sum mode to the oblique difference mode, through the channel $\tau_{wn}$, with $48\,\%$ of the exchange of that triad, in the interval $\xi=40$ to $52$ in which both modes acquire spanwise velocity. This one triad moves $0.4$ of the production of the oblique sum mode. The roller supplies the advecting velocity of the exchange, not its energy. The rollers become three-dimensional through this channel, which starts at the height of the roller cores. The routes examined are confined to a thin sheet at the transition front, spanwise periodic when the flapping mode is the third member and spanwise coherent when the Kelvin-Helmholtz mode is the catalyst.

Three guidelines follow for a reduced-order description of the present bubble. Such a description must retain the flapping mode and the Kelvin-Helmholtz mode, the third members of all the strongest triads. It must retain at least one oblique plane at the roller wavelength, on which the exchange of spanwise velocity occurs. Its closure cannot be low-rank, because the transfer that feeds the tonal modes is spread across several hundred modes and a truncation to the nine tonal modes retains $19\,\%$ of the exchange they take part in.

\begin{bmhead}[Acknowledgements.]
We acknowledge ISCRA for awarding this project access to the LEONARDO supercomputer, owned by the EuroHPC Joint Undertaking and hosted by CINECA (Italy).
\end{bmhead}

\begin{bmhead}[Funding.]
This project has received funding from the European Union's Horizon Europe research and innovation programme under the Marie Sklodowska-Curie grant agreement no.~101205669.
\end{bmhead}

\begin{bmhead}[Declaration of interests.]
The authors report no conflict of interest. The data are available upon reasonable request.
\end{bmhead}

\begin{bmhead}[Author ORCIDs.]
G. Lopes, https://orcid.org/0009-0009-5061-897X; P. Przytarski, https://orcid.org/0000-0003-0104-4606; D. Simoni, https://orcid.org/0000-0003-4161-0721; D. Lengani, https://orcid.org/0000-0001-6347-0817.
\end{bmhead}

\appendix

\section{Mode-sum identities of the spectral balance}\label{app:sum}
The transform (\ref{eq:dft}) is unitary, so for any two fluctuation fields $a'(x,y,z,t)$ and $b'(x,y,z,t)$ with transforms $\hat{a}$, $\hat{b}$, Parseval's identity gives
\begin{equation}
  \sum_{f,\beta}\Re\big[\hat{a}\,\hat{b}^{*}\big]
  =\frac{1}{N N_z}\sum_{q,r} a'b'
  =\overline{a'b'},
  \label{eq:parseval}
\end{equation}
where the overbar is the record and span average that defines the statistics of (\ref{eq:phys}). Every linear term of (\ref{eq:spectral}) is of this form. It is a quadratic correlation whose coefficients, the mean velocity, its gradients, $\nu$ and $1/\rho$, are independent of $t$ and $z$ and pass through the mode sum. Applying (\ref{eq:parseval}) term by term, the mode sums of $\hat{\mathcal{C}}$, $\hat{\mathcal{P}}$, $\hat{\mathcal{D}}_p$, $\hat{\mathcal{D}}_v$ and $\hat{\varepsilon}$ recover $\mathcal{C}$, $\mathcal{P}$, $\mathcal{D}_p$, $\mathcal{D}_v$ and $\varepsilon$. For the production, for instance,
\begin{equation}
  -\sum_{f,\beta}\Re\big[\hat{u}_i\hat{u}_j^{*}\big]
  \frac{\partial U_i}{\partial x_j}
  =-\overline{u_i'u_j'}\,\frac{\partial U_i}{\partial x_j}
  =\mathcal{P}.
\end{equation}

The correspondence of the cubic term is established starting from the transport term of (\ref{eq:phys}). Expanding its divergence by the product rule,
\begin{equation}
  \mathcal{D}_T
  =-\frac{\partial}{\partial x_j}
    \Big(\frac{1}{2}\overline{u_i'u_i'u_j'}\Big)
  =-\,\overline{u_j'\,\frac{\partial}{\partial x_j}
    \Big(\frac{1}{2}u_i'u_i'\Big)}
  -\,\overline{\frac{1}{2}u_i'u_i'\,
    \frac{\partial u_j'}{\partial x_j}},
\end{equation}
and the second term vanishes by the continuity equation $\partial u_j'/\partial x_j=0$. The chain rule on the remaining term, $\frac{\partial}{\partial x_j}(\frac{1}{2}u_i'u_i')=u_i'\,\frac{\partial u_i'}{\partial x_j}$, gives
\begin{equation}
  \mathcal{D}_T
  =-\,\overline{u_i'\,u_j'\,\frac{\partial u_i'}{\partial x_j}}.
\end{equation}
This is a correlation of the form (\ref{eq:parseval}) with $a'=u_i'$ and $b'=u_j'\,\partial u_i'/\partial x_j$, so
\begin{equation}
  \mathcal{D}_T
  =-\sum_{f,\beta}\Re\Big[\hat{u}_i^{*}\,
    \widehat{u_j'\tfrac{\partial u_i'}{\partial x_j}}\,\Big]
  =\sum_{f,\beta}\hat{\mathcal{N}}.
  \label{eq:cubicsum}
\end{equation}
The mode sum of the nonlinear term is therefore exactly the spatial transport of (\ref{eq:phys}). Mode by mode, $\hat{\mathcal{N}}$ additionally redistributes energy between scales. This interscale part carries no mode-summed content. That is the pointwise zero-sum property (\ref{eq:trzero}) and, in the triad-resolved form (\ref{eq:tensor}), the antisymmetric conservation of the transfer \citep{Waleffe1992,Cheung2014}.

\bibliographystyle{jfm}
\bibliography{jfm}

\begin{thebibliography}{54}
\expandafter\ifx\csname natexlab\endcsname\relax\def\natexlab#1{#1}\fi
\def\au#1{#1} \def\ed#1{#1} \def\yr#1{#1}\def\at#1{#1}\def\jt#1{\textit{#1}}
  \def\bt#1{#1}\def\bvol#1{\textbf{#1}} \def\vol#1{#1} \def\pg#1{#1}
  \def\publ#1{#1}\def\arxiv#1{#1}\def\org#1{#1}\def\st#1{\textit{#1}}

\bibitem[Alam \& Sandham(2000)]{Alam2000}
{\sc \au{Alam, M.} \& \au{Sandham, N.~D.}} \yr{2000}  \at{Direct numerical
  simulation of `short' laminar separation bubbles with turbulent
  reattachment}.  \jt{Journal of Fluid Mechanics}  \bvol{410},  \pg{1--28}.

\bibitem[Baj \& Buxton(2017)]{Baj2017}
{\sc \au{Baj, P.} \& \au{Buxton, O. R.~H.}} \yr{2017}  \at{Interscale energy
  transfer in the merger of wakes of a multiscale array of rectangular
  cylinders}.  \jt{Physical Review Fluids}  \bvol{2},  \pg{114607}.

\bibitem[Ben Hassan~Sa{\"\i}di {\em et~al.\/}(2025)Ben Hassan~Sa{\"\i}di, Wang,
  Fournier, Tenaud \& Robinet]{BenHassanSaidi2025}
{\sc \au{Ben Hassan~Sa{\"\i}di, I.}, \au{Wang, S.}, \au{Fournier, G.},
  \au{Tenaud, C.} \& \au{Robinet, J.-C.}} \yr{2025}  \at{Modal analysis of the
  triadic interactions in the dynamics of a transitional shock wave boundary
  layer interaction}.  \jt{Journal of Fluid Mechanics}  \bvol{1009},  \pg{A43}.

\bibitem[Biswas {\em et~al.\/}(2022)Biswas, Cicolin \& Buxton]{Biswas2022}
{\sc \au{Biswas, N.}, \au{Cicolin, M.~M.} \& \au{Buxton, O. R.~H.}} \yr{2022}
  \at{Energy exchanges in the flow past a cylinder with a leeward control rod}.
   \jt{Journal of Fluid Mechanics}  \bvol{941},  \pg{A36}.

\bibitem[Borgmann {\em et~al.\/}(2025)Borgmann, Little \& Fasel]{Borgmann2025}
{\sc \au{Borgmann, D.}, \au{Little, J.} \& \au{Fasel, H.~F.}} \yr{2025}
  \at{Active control of transition to turbulence in laminar separation
  bubbles}.  \jt{Journal of Fluid Mechanics}  \bvol{1016},  \pg{A59}.

\bibitem[Chen \& Garc{\'i}a-Mayoral(2026)]{ChenGarciaMayoral2026}
{\sc \au{Chen, J.} \& \au{Garc{\'i}a-Mayoral, R.}} \yr{2026}  \at{Inter-scale
  energy transfer in turbulent channels}.  \jt{J. Phys.: Conf. Ser.}
  \bvol{3230},  \pg{012007}.

\bibitem[Cheung \& Zaki(2014)]{Cheung2014}
{\sc \au{Cheung, L.~C.} \& \au{Zaki, T.~A.}} \yr{2014}  \at{An exact
  representation of the nonlinear triad interaction terms in spectral space}.
  \jt{Journal of Fluid Mechanics}  \bvol{748},  \pg{175--188}.

\bibitem[Cho {\em et~al.\/}(2018)Cho, Hwang \& Choi]{Cho2018}
{\sc \au{Cho, M.}, \au{Hwang, Y.} \& \au{Choi, H.}} \yr{2018}  \at{Scale
  interactions and spectral energy transfer in turbulent channel flow}.
  \jt{Journal of Fluid Mechanics}  \bvol{854},  \pg{474--504}.

\bibitem[Cimarelli {\em et~al.\/}(2019)Cimarelli, Leonforte, De~Angelis,
  Crivellini \& Angeli]{Cimarelli2019}
{\sc \au{Cimarelli, A.}, \au{Leonforte, A.}, \au{De~Angelis, E.},
  \au{Crivellini, A.} \& \au{Angeli, D.}} \yr{2019}  \at{On negative turbulence
  production phenomena in the shear layer of separating and reattaching flows}.
   \jt{Physics Letters A}  \bvol{383}~(10),  \pg{1019--1026}.

\bibitem[Coleman {\em et~al.\/}(2018)Coleman, Rumsey \& Spalart]{Coleman2018}
{\sc \au{Coleman, G.~N.}, \au{Rumsey, C.~L.} \& \au{Spalart, P.~R.}} \yr{2018}
  \at{Numerical study of turbulent separation bubbles with varying pressure
  gradient and {R}eynolds number}.  \jt{Journal of Fluid Mechanics}
  \bvol{847},  \pg{28--70}.

\bibitem[Craik(1971)]{Craik1971}
{\sc \au{Craik, A. D.~D.}} \yr{1971}  \at{Non-linear resonant instability in
  boundary layers}.  \jt{Journal of Fluid Mechanics}  \bvol{50}~(2),
  \pg{393--413}.

\bibitem[Dellacasagrande {\em et~al.\/}(2023)Dellacasagrande, Lengani, Simoni
  \& Yarusevych]{Dellacasagrande2023}
{\sc \au{Dellacasagrande, M.}, \au{Lengani, D.}, \au{Simoni, D.} \&
  \au{Yarusevych, S.}} \yr{2023}  \at{A data-driven analysis of short and long
  laminar separation bubbles}.  \jt{Journal of Fluid Mechanics}  \bvol{976},
  \pg{R3}.

\bibitem[Deuse \& Sandberg(2020)]{Deuse2020}
{\sc \au{Deuse, M.} \& \au{Sandberg, R.~D.}} \yr{2020}  \at{Implementation of a
  stable high-order overset grid method for high-fidelity simulations}.
  \jt{Computers \& Fluids}  \bvol{211},  \pg{104449}.

\bibitem[Ding {\em et~al.\/}(2025)Ding, Chung \& Illingworth]{Ding2025}
{\sc \au{Ding, J.}, \au{Chung, D.} \& \au{Illingworth, S.~J.}} \yr{2025}
  \at{Mode-to-mode nonlinear energy transfer in turbulent channel flows}.
  \jt{Journal of Fluid Mechanics}  \bvol{1002},  \pg{A42}.

\bibitem[Diwan {\em et~al.\/}(2006)Diwan, Chetan \& Ramesh]{Diwan2006}
{\sc \au{Diwan, S.~S.}, \au{Chetan, S.~J.} \& \au{Ramesh, O.~N.}} \yr{2006}
  \at{On the bursting criterion for laminar separation bubbles}.  \bt{In {\em
  IUTAM Symposium on Laminar-Turbulent Transition\/} (ed.
  \ed{R.~Govindarajan})},  \st{Fluid Mechanics and Its Applications},
  \vol{vol.~78},  \pg{pp. 401--407}.  \publ{Springer}.

\bibitem[Embacher \& Fasel(2014)]{Embacher2014}
{\sc \au{Embacher, M.} \& \au{Fasel, H.~F.}} \yr{2014}  \at{Direct numerical
  simulations of laminar separation bubbles: investigation of absolute
  instability and active flow control of transition to turbulence}.
  \jt{Journal of Fluid Mechanics}  \bvol{747},  \pg{141--185}.

\bibitem[Fang \& Wang(2024)]{Fang2024}
{\sc \au{Fang, X.} \& \au{Wang, Z.}} \yr{2024}  \at{On the low-frequency
  flapping motion in flow separation}.  \jt{Journal of Fluid Mechanics}
  \bvol{984},  \pg{A76}.

\bibitem[Gaster(1967)]{Gaster1967}
{\sc \au{Gaster, M.}} \yr{1967}  \bt{The structure and behaviour of separation
  bubbles}. Reports and Memoranda 3595.  \org{Aeronautical Research Council}.

\bibitem[Ho \& Huerre(1984)]{HoHuerre1984}
{\sc \au{Ho, C.-M.} \& \au{Huerre, P.}} \yr{1984}  \at{Perturbed free shear
  layers}.  \jt{Annual Review of Fluid Mechanics}  \bvol{16},  \pg{365--424}.

\bibitem[Hosseinverdi \& Fasel(2019)]{Hosseinverdi2019}
{\sc \au{Hosseinverdi, S.} \& \au{Fasel, H.~F.}} \yr{2019}  \at{Numerical
  investigation of laminar--turbulent transition in laminar separation bubbles:
  the effect of free-stream turbulence}.  \jt{Journal of Fluid Mechanics}
  \bvol{858},  \pg{714--759}.

\bibitem[Jones {\em et~al.\/}(2008)Jones, Sandberg \& Sandham]{Jones2008}
{\sc \au{Jones, L.~E.}, \au{Sandberg, R.~D.} \& \au{Sandham, N.~D.}} \yr{2008}
  \at{Direct numerical simulations of forced and unforced separation bubbles on
  an airfoil at incidence}.  \jt{Journal of Fluid Mechanics}  \bvol{602},
  \pg{175--207}.

\bibitem[Kawata \& Alfredsson(2018)]{Kawata2018}
{\sc \au{Kawata, T.} \& \au{Alfredsson, P.~H.}} \yr{2018}  \at{Inverse
  interscale transport of the reynolds shear stress in plane couette
  turbulence}.  \jt{Physical Review Letters}  \bvol{120},  \pg{244501}.

\bibitem[Kinjangi \& Foti(2023)]{Kinjangi2023}
{\sc \au{Kinjangi, D.~K.} \& \au{Foti, D.}} \yr{2023}  \at{Characterization of
  energy transfer and triadic interactions of coherent structures in turbulent
  wakes}.  \jt{Journal of Fluid Mechanics}  \bvol{971},  \pg{A7}.

\bibitem[Kurelek {\em et~al.\/}(2018)Kurelek, Kotsonis \&
  Yarusevych]{Kurelek2018}
{\sc \au{Kurelek, J.~W.}, \au{Kotsonis, M.} \& \au{Yarusevych, S.}} \yr{2018}
  \at{Transition in a separation bubble under tonal and broadband acoustic
  excitation}.  \jt{Journal of Fluid Mechanics}  \bvol{853},  \pg{1--36}.

\bibitem[Kurelek {\em et~al.\/}(2019)Kurelek, Yarusevych \&
  Kotsonis]{Kurelek2019}
{\sc \au{Kurelek, J.~W.}, \au{Yarusevych, S.} \& \au{Kotsonis, M.}} \yr{2019}
  \at{Vortex merging in a laminar separation bubble under natural and forced
  conditions}.  \jt{Physical Review Fluids}  \bvol{4}~(6),  \pg{063903}.

\bibitem[Lee \& Moser(2019)]{Lee2019}
{\sc \au{Lee, M.} \& \au{Moser, R.~D.}} \yr{2019}  \at{Spectral analysis of the
  budget equation in turbulent channel flows at high reynolds number}.
  \jt{Journal of Fluid Mechanics}  \bvol{860},  \pg{886--938}.

\bibitem[Lengani {\em et~al.\/}(2017)Lengani, Simoni, Ubaldi, Zunino \&
  Bertini]{Lengani2017}
{\sc \au{Lengani, D.}, \au{Simoni, D.}, \au{Ubaldi, M.}, \au{Zunino, P.} \&
  \au{Bertini, F.}} \yr{2017}  \at{Analysis of the {R}eynolds stress component
  production in a laminar separation bubble}.  \jt{International Journal of
  Heat and Fluid Flow}  \bvol{64},  \pg{112--119}.

\bibitem[Lopes {\em et~al.\/}(2026)Lopes, Rosenzweig, Przytarski, Sandberg \&
  Lengani]{Lopes2026}
{\sc \au{Lopes, G.}, \au{Rosenzweig, M.}, \au{Przytarski, P.~J.}, \au{Sandberg,
  R.~D.} \& \au{Lengani, D.}} \yr{2026}  \at{A {GPU} framework for computing
  multiscale triadic interactions in turbulent flows}.  \jt{Computers \&
  Fluids}  \pg{p. 107298}.

\bibitem[Mansour {\em et~al.\/}(1988)Mansour, Kim \& Moin]{Mansour1988}
{\sc \au{Mansour, N.~N.}, \au{Kim, J.} \& \au{Moin, P.}} \yr{1988}
  \at{Reynolds-stress and dissipation-rate budgets in a turbulent channel
  flow}.  \jt{Journal of Fluid Mechanics}  \bvol{194},  \pg{15--44}.

\bibitem[Marxen {\em et~al.\/}(2013)Marxen, Lang \& Rist]{Marxen2013}
{\sc \au{Marxen, O.}, \au{Lang, M.} \& \au{Rist, U.}} \yr{2013}  \at{Vortex
  formation and vortex breakup in a laminar separation bubble}.  \jt{Journal of
  Fluid Mechanics}  \bvol{728},  \pg{58--90}.

\bibitem[Michelis {\em et~al.\/}(2018)Michelis, Yarusevych \&
  Kotsonis]{Michelis2018}
{\sc \au{Michelis, T.}, \au{Yarusevych, S.} \& \au{Kotsonis, M.}} \yr{2018}
  \at{On the origin of spanwise vortex deformations in laminar separation
  bubbles}.  \jt{Journal of Fluid Mechanics}  \bvol{841},  \pg{81--108}.

\bibitem[Miksad(1973)]{Miksad1973}
{\sc \au{Miksad, R.~W.}} \yr{1973}  \at{Experiments on nonlinear interactions
  in the transition of a free shear layer}.  \jt{Journal of Fluid Mechanics}
  \bvol{59}~(1),  \pg{1--21}.

\bibitem[Mizuno(2016)]{Mizuno2016}
{\sc \au{Mizuno, Y.}} \yr{2016}  \at{Spectra of energy transport in turbulent
  channel flows for moderate reynolds numbers}.  \jt{Journal of Fluid
  Mechanics}  \bvol{805},  \pg{171--187}.

\bibitem[Moffatt(2014)]{Moffatt2014}
{\sc \au{Moffatt, H.~K.}} \yr{2014}  \at{Note on the triad interactions of
  homogeneous turbulence}.  \jt{Journal of Fluid Mechanics}  \bvol{741},
  \pg{R3}.

\bibitem[Moniripiri {\em et~al.\/}(2025)Moniripiri, Rodr{\'i}guez \&
  Hanifi]{Moniripiri2025}
{\sc \au{Moniripiri, M.}, \au{Rodr{\'i}guez, D.} \& \au{Hanifi, A.}} \yr{2025}
  \at{Self-excited instability and transition to turbulence in laminar
  separation bubbles induced by surface waviness}.  \jt{J. Fluid Mech.}
  \bvol{1011},  \pg{A27}.

\bibitem[Nekkanti {\em et~al.\/}(2025{\natexlab{{\em a\/}}})Nekkanti, Colonius
  \& Schmidt]{Nekkanti2025b}
{\sc \au{Nekkanti, A.}, \au{Colonius, T.} \& \au{Schmidt, O.~T.}}
  \yr{2025{\natexlab{{\em a\/}}}}  \at{Nonlinear dynamics of vortex pairing in
  transitional jets}.  \jt{Journal of Fluid Mechanics}  \bvol{1018},  \pg{A32}.

\bibitem[Nekkanti {\em et~al.\/}(2025{\natexlab{{\em b\/}}})Nekkanti,
  Pickering, Schmidt \& Colonius]{Nekkanti2025}
{\sc \au{Nekkanti, A.}, \au{Pickering, E.}, \au{Schmidt, O.~T.} \&
  \au{Colonius, T.}} \yr{2025{\natexlab{{\em b\/}}}}  \at{Bispectral
  decomposition and energy transfer in a turbulent jet}.  \jt{Journal of Fluid
  Mechanics}  \bvol{1025},  \pg{A35}.

\bibitem[O'Meara \& Mueller(1987)]{OMeara1987}
{\sc \au{O'Meara, M.~M.} \& \au{Mueller, T.~J.}} \yr{1987}  \at{Laminar
  separation bubble characteristics on an airfoil at low reynolds numbers}.
  \jt{AIAA Journal}  \bvol{25}~(8),  \pg{1033--1041}.

\bibitem[Owen \& Klanfer(1953)]{OwenKlanfer1953}
{\sc \au{Owen, P.~R.} \& \au{Klanfer, L.}} \yr{1953}  \bt{On the laminar
  boundary layer separation from the leading edge of a thin aerofoil}. RAE
  Report Aero. 2508.  \org{Royal Aircraft Establishment}, reissued as A.R.C.
  C.P. 220 (1955).

\bibitem[Przytarski {\em et~al.\/}(2024)Przytarski, Dellacasagrande, Kozul,
  Agrawal \& Bose]{PrzytarskiCTR2024}
{\sc \au{Przytarski, P.~J.}, \au{Dellacasagrande, M.}, \au{Kozul, M.},
  \au{Agrawal, R.} \& \au{Bose, S.~T.}} \yr{2024} Multiscale interactions
  within a multistage compressor.  \bt{In {\em Proceedings of the Summer
  Program 2024\/}}.  \publ{Center for Turbulence Research, Stanford
  University}.

\bibitem[Przytarski \& Wheeler(2021)]{Przytarski2021}
{\sc \au{Przytarski, P.~J.} \& \au{Wheeler, A. P.~S.}} \yr{2021}  \at{Accurate
  prediction of loss using high fidelity methods}.  \jt{Journal of
  Turbomachinery}  \bvol{143}~(3),  \pg{031008}.

\bibitem[Rinoie \& Takemura(2004)]{Rinoie2004}
{\sc \au{Rinoie, K.} \& \au{Takemura, N.}} \yr{2004}  \at{Oscillating behaviour
  of laminar separation bubble formed on an aerofoil near stall}.  \jt{The
  Aeronautical Journal}  \bvol{108}~(1081),  \pg{153--163}.

\bibitem[Rist \& Maucher(2002)]{Rist2002}
{\sc \au{Rist, U.} \& \au{Maucher, U.}} \yr{2002}  \at{Investigations of
  time-growing instabilities in laminar separation bubbles}.  \jt{Eur. J. Mech.
  B/Fluids}  \bvol{21},  \pg{495--509}.

\bibitem[Rodr{\'i}guez {\em et~al.\/}(2021)Rodr{\'i}guez, Gennaro \&
  Souza]{Rodriguez2021}
{\sc \au{Rodr{\'i}guez, D.}, \au{Gennaro, E.~M.} \& \au{Souza, L.~F.}}
  \yr{2021}  \at{Self-excited primary and secondary instability of laminar
  separation bubbles}.  \jt{Journal of Fluid Mechanics}  \bvol{906},  \pg{A13}.

\bibitem[Rodr{\'\i}guez \& Theofilis(2010)]{Rodriguez2010}
{\sc \au{Rodr{\'\i}guez, D.} \& \au{Theofilis, V.}} \yr{2010}  \at{Structural
  changes of laminar separation bubbles induced by global linear instability}.
  \jt{Journal of Fluid Mechanics}  \bvol{655},  \pg{280--305}.

\bibitem[de~Salis~Young {\em et~al.\/}(2024)de~Salis~Young, Hao \&
  Garc{\'i}a-Mayoral]{DeSalisYoung2024}
{\sc \au{de~Salis~Young, J.}, \au{Hao, Z.} \& \au{Garc{\'i}a-Mayoral, R.}}
  \yr{2024} Inter-scale causality relations in wall turbulence.  \bt{In {\em
  Proceedings of the 5th Madrid Turbulence Workshop\/}}.

\bibitem[Sandberg {\em et~al.\/}(2015)Sandberg, Michelassi, Pichler, Chen \&
  Johnstone]{Sandberg2015}
{\sc \au{Sandberg, R.~D.}, \au{Michelassi, V.}, \au{Pichler, R.}, \au{Chen, L.}
  \& \au{Johnstone, R.}} \yr{2015}  \at{Compressible direct numerical
  simulation of low-pressure turbines---{P}art {I}: Methodology}.  \jt{Journal
  of Turbomachinery}  \bvol{137}~(5),  \pg{051011}.

\bibitem[Schmidt(2020)]{Schmidt2020}
{\sc \au{Schmidt, O.~T.}} \yr{2020}  \at{Bispectral mode decomposition of
  nonlinear flows}.  \jt{Nonlinear Dynamics}  \bvol{102}~(4),  \pg{2479--2501}.

\bibitem[Steinfurth {\em et~al.\/}(2025)Steinfurth, Li, Scarano \&
  Weiss]{Steinfurth2025}
{\sc \au{Steinfurth, B.}, \au{Li, M.}, \au{Scarano, F.} \& \au{Weiss, J.}}
  \yr{2025}  \at{Three-dimensional low-frequency dynamics of a turbulent
  separation bubble}.  \jt{Journal of Fluid Mechanics}  \bvol{1019},  \pg{R4}.

\bibitem[Waleffe(1992)]{Waleffe1992}
{\sc \au{Waleffe, F.}} \yr{1992}  \at{The nature of triad interactions in
  homogeneous turbulence}.  \jt{Physics of Fluids A}  \bvol{4}~(2),
  \pg{350--363}.

\bibitem[Watmuff(1999)]{Watmuff1999}
{\sc \au{Watmuff, J.~H.}} \yr{1999}  \at{Evolution of a wave packet into vortex
  loops in a laminar separation bubble}.  \jt{Journal of Fluid Mechanics}
  \bvol{397},  \pg{119--169}.

\bibitem[Yao {\em et~al.\/}(2022)Yao, Mollicone \& Papadakis]{Yao2022}
{\sc \au{Yao, H.}, \au{Mollicone, J.-P.} \& \au{Papadakis, G.}} \yr{2022}
  \at{Analysis of interscale energy transfer in a boundary layer undergoing
  bypass transition}.  \jt{Journal of Fluid Mechanics}  \bvol{941},  \pg{A14}.

\bibitem[Yeh {\em et~al.\/}(2020)Yeh, Benton, Taira \& Garmann]{Yeh2020}
{\sc \au{Yeh, C.-A.}, \au{Benton, S.~I.}, \au{Taira, K.} \& \au{Garmann,
  D.~J.}} \yr{2020}  \at{Resolvent analysis of an airfoil laminar separation
  bubble at {$Re=500\,000$}}.  \jt{Physical Review Fluids}  \bvol{5},
  \pg{083906}.

\bibitem[Yeung {\em et~al.\/}(2026)Yeung, Chu \& Schmidt]{Yeung2026}
{\sc \au{Yeung, B.}, \au{Chu, T.} \& \au{Schmidt, O.~T.}} \yr{2026}
  \at{Triadic orthogonal decomposition reveals nonlinearity in fluid flows}.
  \jt{Journal of Fluid Mechanics}  \bvol{1031},  \pg{A34}.

\end{thebibliography}

\end{document}